\documentclass[useAMS,usenatbib]{mn2e}

\usepackage{tabularx}
\usepackage{graphicx}
\usepackage[section]{placeins}
\usepackage{epstopdf}
\usepackage{array}
\usepackage{color}
\usepackage{upgreek}
\usepackage{nicefrac}
\usepackage{float}
\usepackage[figuresright]{rotating}
\usepackage{amsmath}
\usepackage{amssymb}

\def \um {$\upmu$m}

\def \mnras {MNRAS}
\def \aap {A\&A}
\def \araa {ARA\&A}
\def \aj {AJ}
\def \pasp {PASP}
\def \apj {ApJ}
\def \apjs {ApJS}

\title[Dense cores in L1495]{The JCMT and Herschel Gould Belt Surveys: A comparison of SCUBA-2 and Herschel data of dense cores in the Taurus dark cloud L1495}

\author[D. Ward-Thompson et al.]{\parbox{\textwidth}{D. Ward-Thompson$^1$, K. Pattle$^1$, J. M. Kirk$^1$, K. Marsh$^2$, J. Buckle$^{3,4}$, J. Hatchell$^5$, D. J. Nutter$^2$, M. J. Griffin$^2$, J. Di Francesco$^{6,7}$, P. Andr\'e$^8$, 
S. Beaulieu$^9$, D. Berry$^{10}$, H. Broekhoven-Fiene$^7$, M. Currie$^{10}$, M. Fich$^{9}$, T. Jenness$^{11}$, D. Johnstone$^{6,7}$, H. Kirk$^6$, J. Mottram$^{12,13}$, J. Pineda$^{14}$, C. Quinn$^2$, S. Sadavoy$^{13}$, C. Salji$^{3,4}$, S. Tisi$^{9}$, S. Walker-Smith$^{3,4}$, G. White$^{15,16}$, 
T. Hill$^{17}$, V. K\"{o}nyves$^8$, P. Palmeirim$^8$, S. Pezzuto$^{18}$ 
}
\vspace{0.1cm} \\
Affiliations can be found after the references.
}

\begin{document}

\date{}

\pagerange{1--19} 

\pubyear{2016}


\maketitle

\begin{abstract}
We present a comparison of SCUBA-2 850-\um\ and \emph{Herschel} 70--500-\um\ observations of the L1495 filament in the Taurus Molecular Cloud with the goal of characterising the SCUBA-2 Gould Belt Survey (GBS) data set.  We identify and characterise starless cores in three data sets: SCUBA-2 850-\um, \emph{Herschel} 250-\um, and \emph{Herschel} 250-\um\ spatially filtered to mimic the SCUBA-2 data.  SCUBA-2 detects only the highest-surface-brightness sources, principally detecting protostellar sources and starless cores embedded in filaments, while \emph{Herschel} is sensitive to most of the cloud structure, including extended low-surface-brightness emission.  \emph{Herschel} detects considerably more sources than SCUBA-2 even after spatial filtering.  We investigate which properties of a starless core detected by \emph{Herschel} determine its detectability by SCUBA-2, and find that they are the core's temperature and column density (for given dust properties).  For similar-temperature cores, such as those seen in L1495, the surface brightnesses of the cores are determined by their column densities, with the highest-column-density cores being detected by SCUBA-2.  For roughly spherical geometries, column density corresponds to volume density, and so SCUBA-2 selects the densest cores from a population at a given temperature.  This selection effect, which we quantify as a function of distance, makes SCUBA-2 ideal for identifying those cores in \emph{Herschel} catalogues that are closest to forming stars.  Our results can now be used by anyone wishing to use the SCUBA-2 GBS data set.
\end{abstract}


\section{Introduction}

Stars form in dense cores within molecular clouds (e.g. \citealt{strom1975}; \citealt{wilking1989}; \citealt{wardthompson1989}; \citealt{ballesterosparedes2003}). Recent work has indicated that cores form preferentially in filaments within molecular clouds (e.g. \citealt{andre2010}; \citealt{molinari2010}; \citealt{andre2014}). Those cores that do not contain proto-stars are known generically as starless cores \citep{beichman1986}. When starless cores become dense enough to be gravitationally bound they are known as pre-stellar cores (\citealt{wardthompson1994}; \citealt{difrancesco2007}; \citealt{wardthompson2007a}). They then collapse to form Class~0 \citep{andre1993} and then Class~I (\citealt{lada1987}; \citealt{wilking1989}) protostars, which can be seen as point-like sources in the near- and mid-infrared (e.g. \citealt{andre2000}).

\begin{figure*} 
\centering
\includegraphics[width=\textwidth]{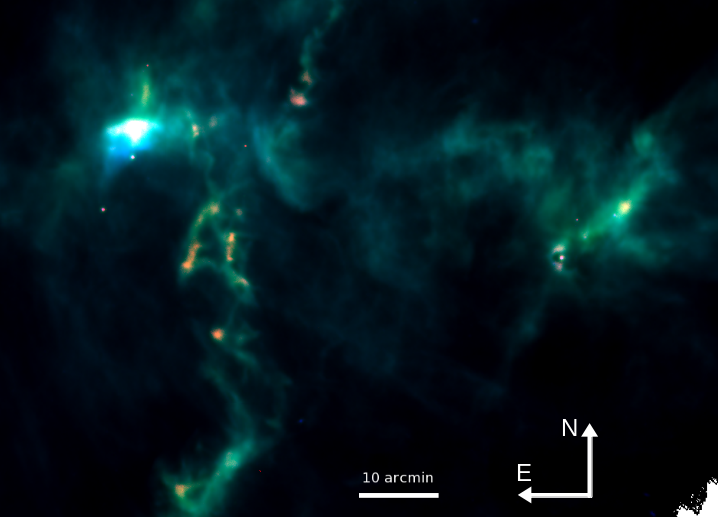}
\caption{The L1495 region, as observed using SCUBA-2 and \emph{Herschel}.  Red channel: SCUBA-2 850-\um\ emission.  Green channel: SPIRE 500-\um\ emission.  Blue channel: SPIRE 250-\um\ emission. Note how most of the structure of the cloud is detected by \emph{Herschel} and is seen as blue-green emission on this image. However, only some of the sources are picked out by SCUBA-2, and show up as red in this image.}
\label{fig:rgb}
\end{figure*}

To trace the structure within molecular clouds, and particularly the filamentary structure, it is necessary to use the far-infrared (e.g. \citealt{andre2010}). The ESA \emph{Herschel} Space Observatory\footnote{\emph{Herschel} is an ESA space observatory with science instruments provided by European-led Principal Investigator consortia and with participation from NASA.} \citep{pilbratt2010} has been especially successful in this regard (e.g. \citealt{andre2010}; \citealt{molinari2010}). In particular the large collecting area and powerful science payload of \emph{Herschel} allows one to perform high-resolution, sensitive, mid- and far-infrared imaging photometry using the PACS \citep{poglitsch2010} and SPIRE \citep{griffin2010} instruments -- often simultaneously (e.g. \citealt{andre2010}).

However, \emph{Herschel} is so sensitive to molecular cloud structure -- particularly SPIRE -- that it sees all of the cores within a cloud (e.g. \citealt{menshchikov2010}). It detects many starless cores that are not pre-stellar in nature (e.g. \citealt{wardthompson2010}), as well as those that are (e.g. \citealt{andre2010}). Consequently, if one is to confirm which cores are gravitationally bound (pre-stellar), and hence destined to form stars, additional information is needed.  The minimum information required in order to determine the virial balance of a starless core is a measure of the core's mass and size (obtainable from submillimetre continuum observations) and internal velocity dispersion (requiring measurement of the core's internal linewidth).  In order to accurately determine both the virial boundedness of the core and the mechanism by which it is confined (i.e. by self-gravity or by external pressure), measures of the external pressure on and magnetic field strength within the core are also required (c.f. \citealt{pattle2015}). 

\begin{figure*} 
\centering
\includegraphics[width=0.8\textwidth]{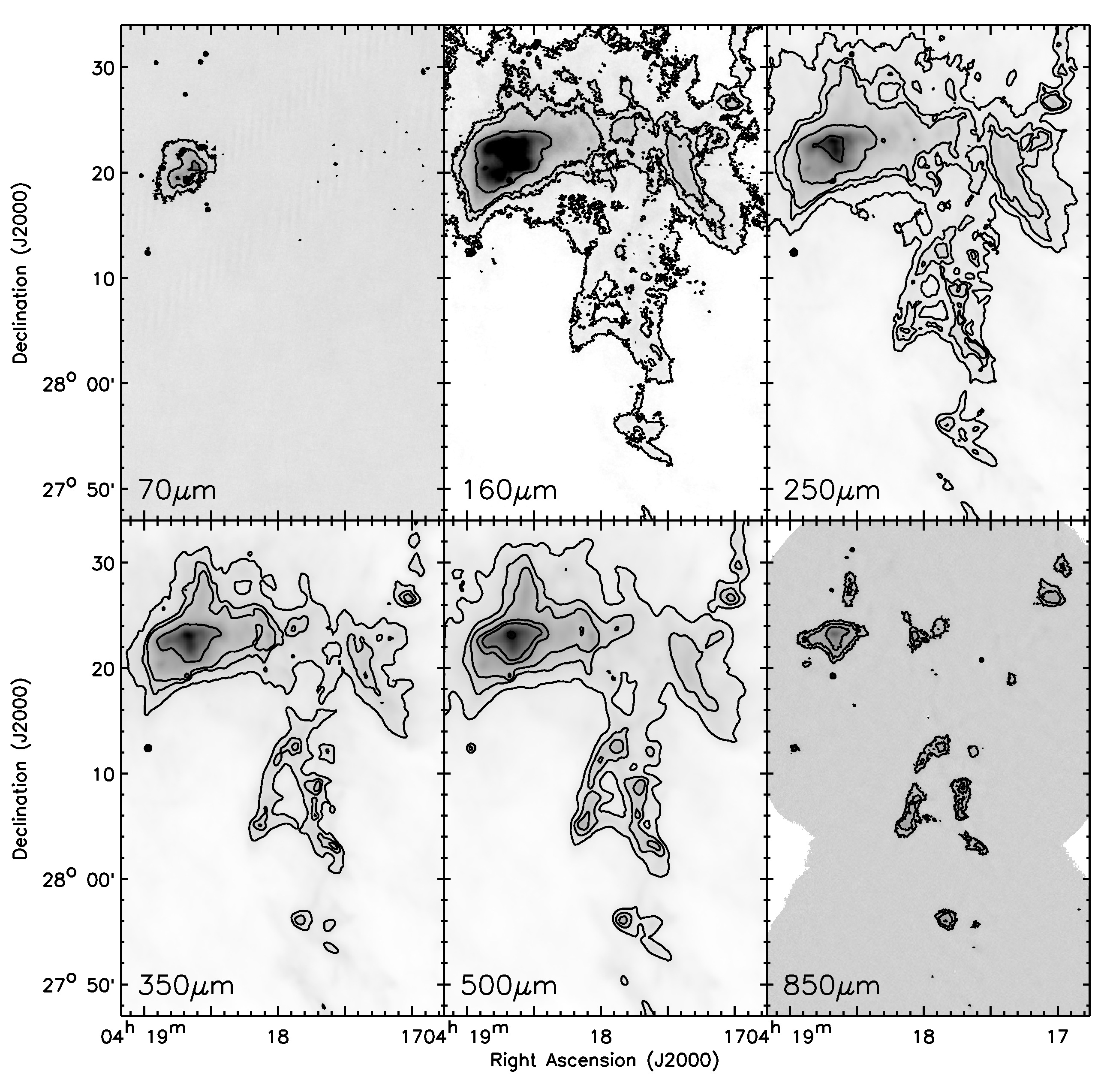}
\caption{The centre of the L1495 region, as observed at the five \emph{Herschel} wavelengths, 70, 160, 250, 350 and 500\um\ \citep{marsh2014,marsh2016} and SCUBA-2 850~\um\ \citep{buckle2015}. Note how only some of the sources and structures seen at other wavelengths are detected by SCUBA-2 (lower right).  Contour levels -- 70\um: 0.01, 0.02, 0.05, 0.5 Jy/6-arcsec pixel; 160\um: 0.05, 0.08, 0.1, 0.2, 0.5, 1.0 Jy/6-arcsec pixel; 250\um: as 160\um; 500\um: 0.02, 0.035, 0.05, 0.08, 0.1, 0.2, 0.5 Jy/6-arcsec pixel; 850\um: 0.005, 0.01, 0.02 Jy/6-arcsec pixel.}
\label{fig:compare}
\end{figure*}

Submillimetre continuum mapping identified the first genuine pre-stellar cores \citep{wardthompson1994}. Therefore, in this paper we test the hypothesis that submillimetre continuum mapping with the Submillimetre Common User Bolometer Array~2 (SCUBA-2) camera can be used to discriminate between those cores detected by \emph{Herschel} which are most likely to be gravitationally bound (pre-stellar), and those which are unbound and transient in nature.  In the absence of the spectroscopic data required to perform a virial analysis, the stability of starless cores is often assessed using density-based criteria such as the Jeans criterion \citet{jeans1928} or the Bonnor-Ebert criterion (\citealt{ebert1955}; \citealt{bonnor1956}).  In these analyses, denser cores are, for a given temperature, more likely to be gravitationally bound.

The presence of the atmosphere means that ground-based submillimetre continuum mapping instruments must inevitably be subject to limitations on absolute sensitivity and large-scale structure recovery to which space-based instruments are not (e.g. \citealt{sadavoy2013}).  However, the selection effects introduced by these constraints may result in a ground-based detection, or otherwise, of a core detected using space-based instrumentation providing additional information about the properties of that core than the space-based data alone could provide.  In this paper we identify and quantify the selection effects determining the detection, or otherwise, with SCUBA-2 of a starless core detected with \emph{Herschel}.  We further investigate whether these selection effects allow the identification of the densest cores in the \emph{Herschel} data through their detection, or otherwise, by SCUBA-2.

The L1495 region of Taurus appears as an obscuring dark cloud on optical images (\citealt{barnard1907}; \citealt{lynds1962}), with a linear, or filamentary, structure, coming to a head at a small globule that is referred to as L1495A (\citealt{benson1989}; \citealt{lee2001}). \citet{hacar2013} studied the L1495 filament and found it to be consistent with the filamentary star formation model \citep{andre2014} favoured by \emph{Herschel} observations of star-forming regions \citep{andre2010}. \citet{lee2001} found evidence for asymmetric line profiles in the southern part of L1495A, apparently indicating collapse or contraction. However, this is close to the Herbig Ae/Be star, V892~Tau, which is clearly affecting at least the southern part of L1495A, so interpreting asymmetric line profiles is made more complex in this area.  The \emph{Herschel} data of the L1495 filament \citep{marsh2016} and follow-up ammonia data from the Green Bank Telescope \citep{seo2015} identified a number of pre-stellar cores in the L1495 filament.

The SCUBA-2 observations presented here were carried out as part of the Gould Belt Legacy Survey (GBLS) on the James Clerk Maxwell Telescope (JCMT) in Hawaii \citep{wardthompson2007}. The full SCUBA-2 data were presented by \citet{buckle2015}.  The \emph{Herschel} observations presented in this paper were carried out as part of the \emph{Herschel} Gould Belt Survey (HGBS) guaranteed-time key programme \citep{andre2010}. The full \emph{Herschel} data were presented by \citet{marsh2014} and \citet{marsh2016}.  \emph{Herschel} observations of the southern part of the L1495 filament were presented by \citet{palmeirim2013}.  In this paper we present a comparison between the two sets of data -- see Figures~\ref{fig:rgb} and~\ref{fig:compare}.  We use the \emph{Herschel} data \citep{marsh2016} as a comparison data set to determine which sources are detected by SCUBA-2, and what properties are important for a SCUBA-2 detection.

\section{Observations}

\subsection{SCUBA-2}

The SCUBA-2 \citep{holland2013} observations used here form part of the JCMT Gould Belt Legacy Survey (GBLS, \citealt{wardthompson2007}).  The L1495 region of the Taurus molecular cloud was observed with the SCUBA-2 camera in 22 observations taken between October 2011 and July 2013. Continuum observations at 450 and 850~\um\ were made using fully sampled 15-, 30-, and 60-arcmin diameter circular regions (PONG900, 1800 and 3600 mapping modes, \citealt{bintley2014}).  Larger regions were mosaicked with overlapping scans. The final output map is centred at a position of R.A. (2000) 04$^{h}$ 17$^{m}$ 54$^{s}$, Dec. (2000) $+$28$^\circ$ 05$^\prime$ 24$^{\prime\prime}$. These data were presented by \citet{buckle2015}, in which full details of the observations are given.  Here, we briefly reiterate the key points.  We only use the 850-\um\ data\footnote{The SCUBA-2 data used in this paper are available at: http://dx.doi.org/10.11570/16.0002.}.

\begin{figure*} 
  \centering
  \includegraphics[width=\textwidth]{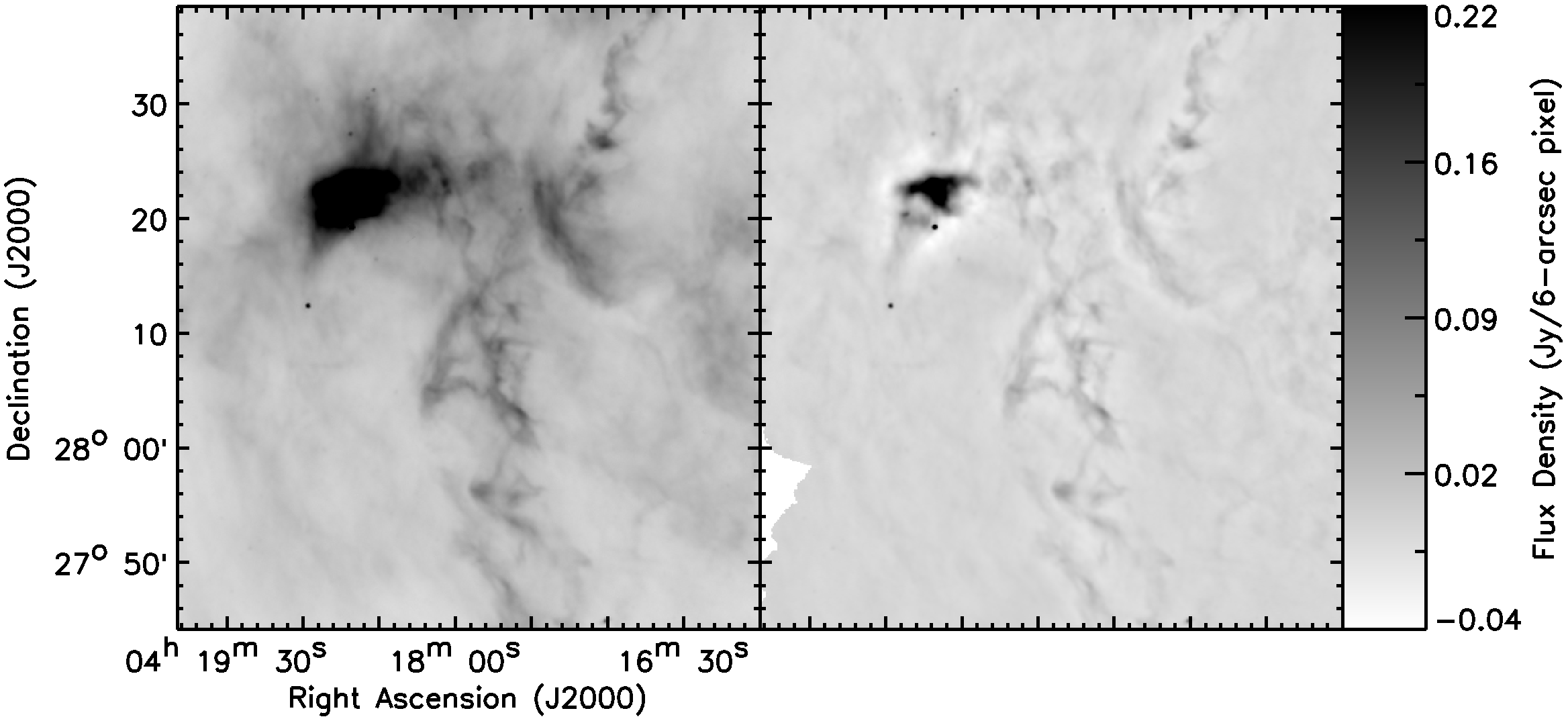}
  \caption{A comparison of the unfiltered (left) and filtered (right) SPIRE 250-\um\ maps of the centre of the L1495 region.  Note the loss of large-scale structure between the unfiltered and the filtered maps.}
  \label{fig:unfilt_filt}
\end{figure*}

The data were reduced as part of the Internal Release 1 data set, using an iterative map-making technique (\emph{makemap} in {\sc smurf}, \citealt{chapin2013}), and gridded to 6-arcsec pixels at 850\um. The iterations were halted when the map pixels, on average, changed by $<$0.1\% of the estimated map rms. The initial reductions of each individual scan were coadded to form a mosaic from which a signal-to-noise mask was produced for each region.  The final mosaic was produced from a second reduction using this mask to define areas of emission. Detection of emission structure and calibration accuracy are uncertain outside of the masked regions. The mask used in the reduction can be seen in the quality array in the reduced datafile (see \citealt{buckle2015}).

A spatial filter of 10 arcmin was used in the reduction, which means that flux recovery is robust for sources with a Gaussian FWHM less than 2.5 arcmin. Sources between 2.5 arcmin and 7.5 arcmin may be detected, but both the flux density and the size will be underestimated because Fourier components with scales greater than 5 arcmin are essentially removed by this filtering process. Detection of sources larger than 7.5 arcmin is dependent on the mask used for each reduction. 

The data were calibrated in Jy/pixel, using an aperture flux conversion factor (FCF) of 2.34~Jy/pW/arcsec$^{2}$ at 850~\um, derived from average values of JCMT calibrators \citep{dempsey2013}, and correcting for the pixel area. The PONG scan pattern leads to lower noise in the map centre and overlap regions, while data reduction and emission artifacts can lead to small variations in the noise over the whole map.

\renewcommand{\thefigure}{\arabic{figure}a} 
\begin{figure}
  \centering
  \includegraphics[width=0.5\textwidth]{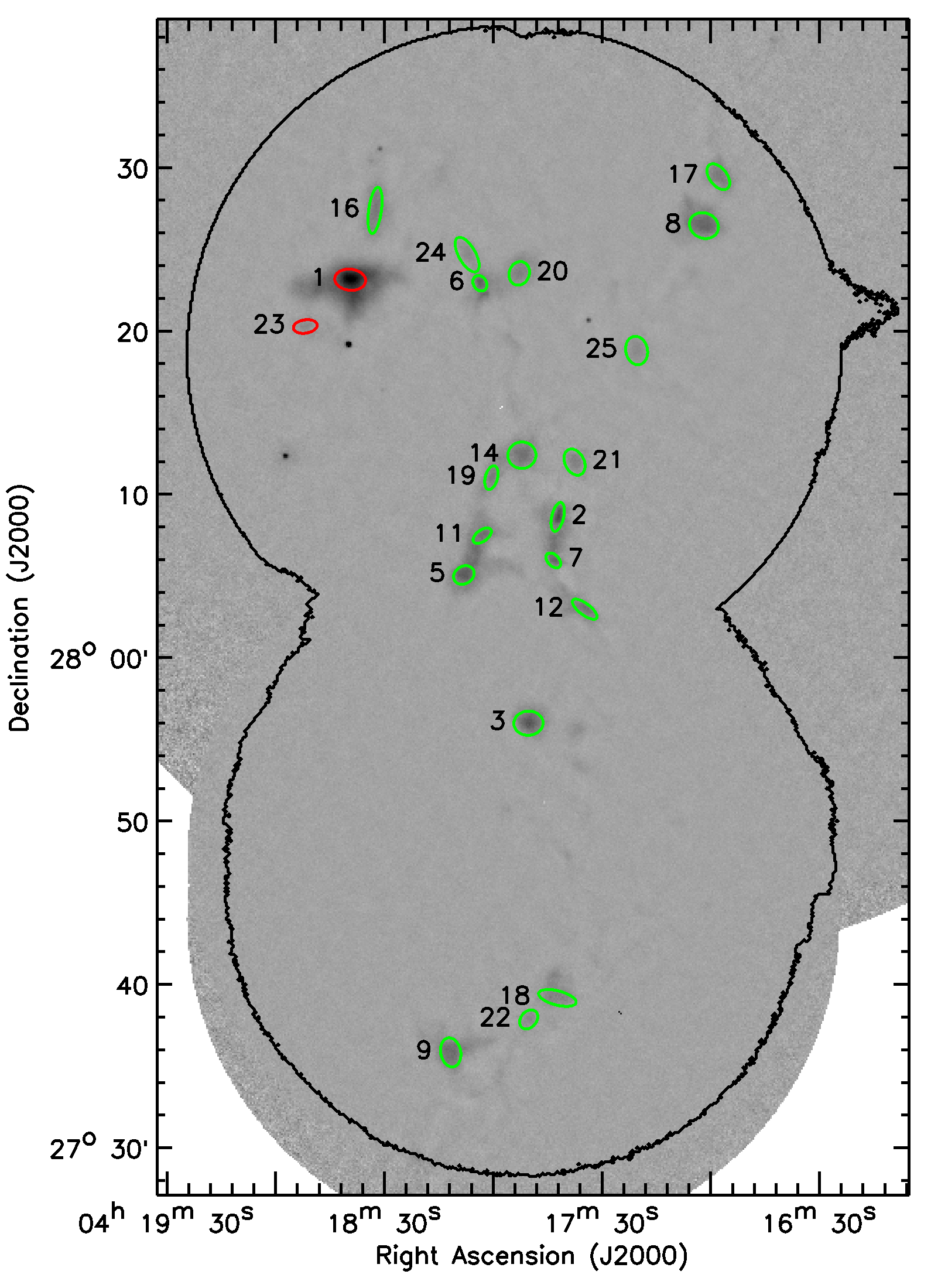}
  \caption{Grey-scale image of the head of the L1495 filament, as mapped by SCUBA-2 at 850~\um. Sources detected in emission by SCUBA-2 are marked by small ellipses.  Cores marked in red show signs of local heating (see text for details). The large-scale contour surrounds the region of lowest variance (c.f. \citealt{buckle2015}). Cores are numbered as in Table~\ref{tab:scuba2_sources}.}
  \label{fig:finding_chart1}
\addtocounter{figure}{-1}
\renewcommand{\thefigure}{\arabic{figure}b}
  \centering
  \includegraphics[width=0.5\textwidth]{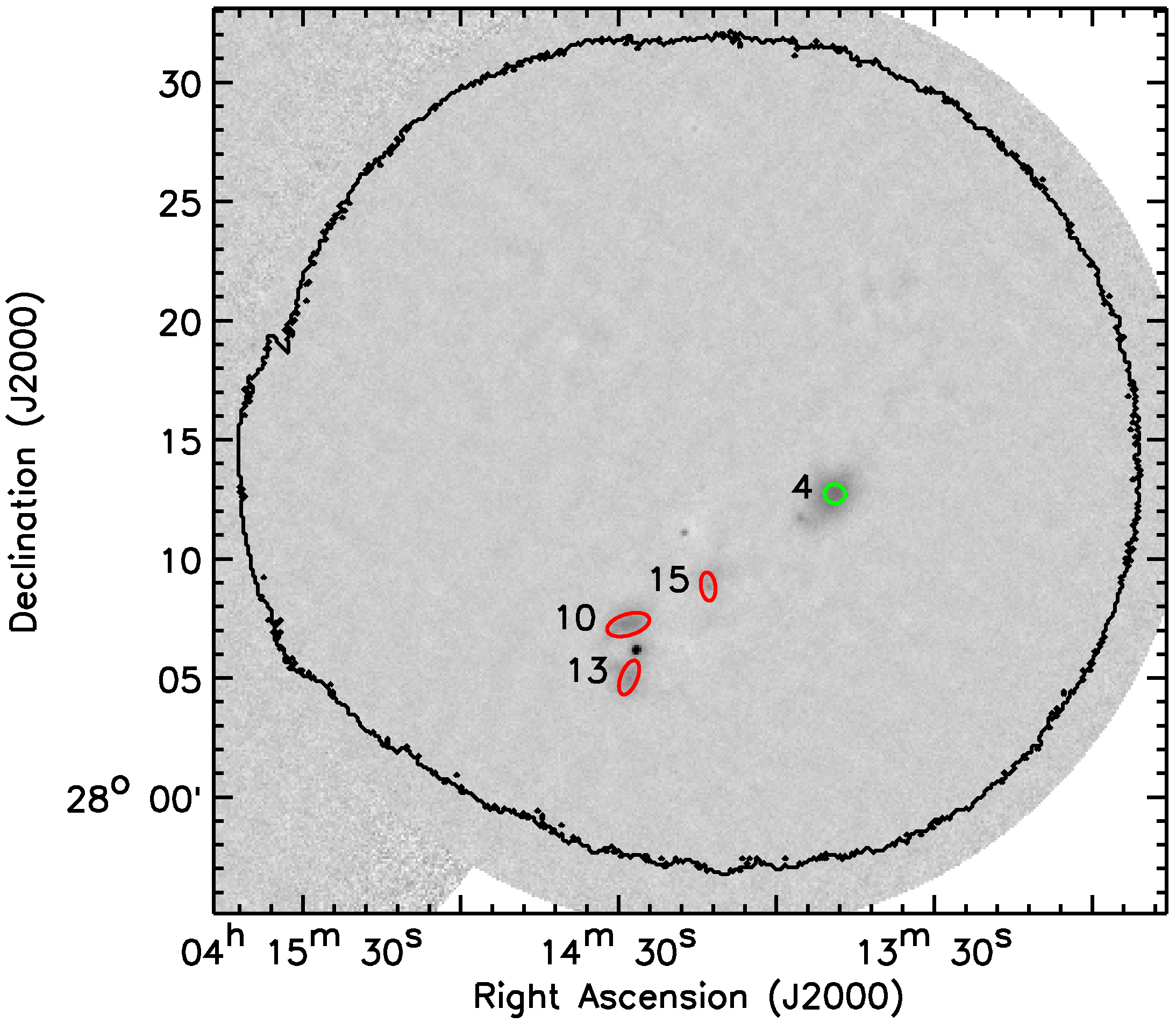}
  \caption{As Figure~\ref{fig:finding_chart1}, for the L1495 West region.}
  \label{fig:finding_chart2}
\end{figure}
\renewcommand{\thefigure}{\arabic{figure}}

\subsection{\emph{Herschel} Space Observatory}

The \emph{Herschel} Space Observatory was a 3.5m-diameter telescope, which operated in the far-infrared and submillimetre regimes \citep{pilbratt2010}.  The comparison \emph{Herschel} data used in this paper were taken as part of the \emph{Herschel} Gould Belt Survey \citep{andre2010} and were presented by \citet{marsh2016}.  They were taken simultaneously with the Photodetector Array Camera and Spectrometer, PACS \citep{poglitsch2010}, and the Spectral and Photometric Imaging Receiver, SPIRE \citep{griffin2010,swinyard2010} using the combined fast-scanning (60\,arcsec/s) SPIRE/PACS parallel mode.  See \citet{andre2010} and \citet{marsh2016} for details of the observations and the data reduction process.

\section{Results}

As SCUBA-2 is a ground-based instrument, while \emph{Herschel} is space-based, the ability of SCUBA-2 to recover submillimetre emission is restricted in comparison to \emph{Herschel}.  Although the data sets used in this work are of similar sensitivity (see below), the SCUBA-2 data are affected by selection effects due to atmospheric emission and variability, to which the \emph{Herschel} data are not.  The result of this is that the emission detected by SCUBA-2 is a subset of that detected by \emph{Herschel}.  We here investigate what distinguishes those sources detected in both SCUBA-2 850-\um\ and \emph{Herschel}-SPIRE 250-\um\ emission from those detected in 250-\um\ emission alone.  We restrict our analysis to extended, starless sources, in order to be able to accurately characterise our sources using only data at wavelengths $>$100\,\um\ and we use a modified blackbody emission model.

The requirement for a SCUBA-2 detection of any source is a peaked 850-\um\ surface brightness, as SCUBA-2 loses sensitivity to flux on larger spatial scales (see above).  Thus, we hypothesise that the likely requirements for a SCUBA-2 detection of a starless core are for the core to be at high density (thus having a high surface brightness), low temperature (i.e. having a high ratio of long-wavelength to short-wavelength flux), and compactness (i.e. being small enough not to lose emission to the SCUBA-2 spatial filtering).  These properties are related to one another: for starless cores of the same mass, in the absence of local heating, a dense core is expected to be colder than a rarefied core; and, trivially, a compact core will be denser than an extended core.  The aim of this study is to distinguish which, if any, of these properties is of most importance in determining whether a starless core identified in \emph{Herschel} data will also be detectable in SCUBA-2 850-\um\ emission.

Figure~\ref{fig:rgb} shows a three-colour image of the region mapped with SCUBA-2, in which 850~\um\ is shown in red, superposed on the same region from the \emph{Herschel} data, where 500~\um\ is shown in green and 250~\um\ is shown in blue. A number of cores and filaments can be seen. \emph{Herschel} detects most of the cloud structure, seen as blue-green emission on this image, including many filaments, as previously seen in other regions. We also see many cores along the filaments, consistent with the recently proposed hypothesis that core formation on filaments is the dominant mode of star formation \citep{andre2014}. However, only some of the sources are picked out by SCUBA-2, and show up as red in this image.

The brightest source in Figure~\ref{fig:rgb} is L1495A. Figure~\ref{fig:compare} shows an enlargement of the L1495A region at each of the 6 wavelengths -- 5 from \emph{Herschel} and one from SCUBA-2. It can be seen that only some of the sources and structures seen at other wavelengths are detected by SCUBA-2. Figures~\ref{fig:finding_chart1} and~\ref{fig:finding_chart2} show the full area covered at high signal-to-noise ratio by SCUBA-2 at 850~\um.

L1495A can be seen clearly at all wavelengths in Figures~\ref{fig:rgb} and~\ref{fig:compare}, including with SCUBA-2 at 850~\um. In fact, the brightest peak coincides with the southern part of L1495A, namely L1495A-S \citep{benson1989}. The northern extension of L1495A, which is very much fainter, is L1495A-N \citep{lee2001}. In the three-colour images in Figure~\ref{fig:rgb}, a colour gradient can be seen from south to north across L1495A, from blue to green. This would tend to indicate a temperature gradient across this core, with the hotter material in the south, consistent with these earlier findings.

There is in fact a bright star, slightly further to the south of L1495A-S, which is V892 Tau (IRAS04155+2812). This has a point source flux density in the IRAS Catalogue of 30 Jy at 12 microns and 100 Jy at 25 microns, which declines slightly to 70 Jy at 60\um, before climbing again to 170 Jy at 100\um. The 100-\um\ flux density almost certainly includes a contribution from L1495A-S, but otherwise, this is a Herbig Ae/Be star, and it is clearly heating L1495A-S, which is otherwise starless. A similar pattern of an externally heated core was modeled in Cepheus by \citet{nutter2009}, based on a combination of SCUBA and Akari data, and in that paper we cautioned that temperature gradients could affect the appearance of cores at long wavelengths such as these.

\section{Comparison of SCUBA-2 data with \emph{Herschel} data}
\label{sec:s2_herschel}

In order to make SCUBA-2 and \emph{Herschel} observations comparable, the large-scale structure must be removed from, and the SCUBA-2 mask must be applied to, the \emph{Herschel} observations.  To this end, the \emph{Herschel} observations were processed through the SCUBA-2 pipeline, following the method described by \citet{sadavoy2013} and \citet{pattle2015}.  The \emph{Herschel} data are scaled such that they represent a small perturbation on the SCUBA-2 850-\um\ flux densities, and are added to the SCUBA-2 bolometer time series.  The data reduction process, as described above, is repeated. The original SCUBA-2 reduction of the data is then subtracted from the \emph{Herschel} $+$ SCUBA-2 map, and the scaling applied to the \emph{Herschel} data is reversed.  This process produces a spatially-filtered and masked version of the original \emph{Herschel} observations, suitable for comparison with the SCUBA-2 data. This process is repeated once for each SCUBA-2 observing position, for which there were corresponding \emph{Herschel} data, and the resulting filtered maps combined to form a mosaic.

The effect of the spatial filtering process on the SPIRE 250-\um\ data set is shown in Figure~\ref{fig:unfilt_filt}.  Figure~\ref{fig:unfilt_filt} shows the loss of large-scale structure caused by the filtering process.

\begin{table*}
\centering
\caption{Sources found by the CSAR algorithm in its non-hierarchical mode in the 850-\um\ SCUBA-2 data -- see text for details.  Sources labelled `S', listed in this table, are detected in SCUBA-2 850\um\ emission. See text for details.}
\label{tab:scuba2_sources}
\begin{tabular}{c c c c c ccccc c}
\hline
Source & RA & Dec & FWHM & Angle & \multicolumn{5}{c}{F$_{\nu}^{total}$ (Jy)} & Counterpart \\ \cline{6-10}
Index & (J2000) & (J2000) & (arcsec$\times$arcsec) & ($^{\circ}$ E of N) & 160\um\ & 250\um\ & 350\um\ & 500\um\ & 850\um\ & Sources \\
\hline
S1 & 4:18:40.00 & +28:23:15.6 & 57.0$\times$39.0 & 84.0 & 30.99 & 45.27 & 32.86 & 18.70 & 5.21 & H1, F1 \\
S2 & 4:17:42.10 & +28:08:44.4 & 54.6$\times$21.4 & 167.0 & 0.74 & 3.15 & 3.80 & 2.79 & 1.08 & H26, F15 \\
S3 & 4:17:50.18 & +27:56:05.5 & 53.4$\times$44.2 & 93.0 & 2.39 & 7.51 & 7.80 & 5.44 & 2.07 & H47, F14 \\
S4 & 4:13:48.01 & +28:12:38.3 & 26.4$\times$24.1 & 74.0 & 1.11 & 3.00 & 3.13 & 2.47 & 0.80 & H6, F9 \\
S5 & 4:18:08.17 & +28:05:10.3 & 39.6$\times$32.0 & 121.0 & 0.96 & 3.69 & 4.30 & 3.43 & 1.40 & H40, F16 \\
S6 & 4:18:03.83 & +28:23:03.5 & 30.0$\times$23.3 & 33.0 & 0.93 & 2.17 & 2.19 & 1.60 & 0.62 & H7, F12 \\
S7 & 4:17:43.31 & +28:06:04.5 & 32.4$\times$20.7 & 45.0 & 0.48 & 1.60 & 1.70 & 1.22 & 0.51 & H28, F17 \\
S8 & 4:17:01.36 & +28:26:36.0 & 54.6$\times$46.0 & 66.0 & 5.28 & 8.93 & 6.69 & 3.99 & 2.29 & H15, F8 \\
S9 & 4:18:11.60 & +27:35:54.4 & 54.0$\times$36.9 & 11.0 & 2.19 & 5.80 & 6.22 & 4.58 & 1.65 & H41, F19 \\
S10 & 4:14:27.63 & +28:07:11.6 & 55.2$\times$26.2 & 106.0 & 5.19 & 8.56 & 6.45 & 3.82 & 1.03 & H8, F4 \\
S11 & 4:18:03.08 & +28:07:35.2 & 39.0$\times$20.5 & 126.0 & 0.30 & 1.28 & 1.57 & 1.24 & 0.69 & None \\
S12 & 4:17:34.58 & +28:03:05.0 & 55.2$\times$20.3 & 53.0 & 1.98 & 4.39 & 3.85 & 2.51 & 0.74 & H19, F10 \\
S13 & 4:14:27.53 & +28:04:58.1 & 45.6$\times$20.8 & 158.0 & 3.20 & 4.76 & 3.28 & 1.89 & 0.64 & H11, F5 \\
S14 & 4:17:52.08 & +28:12:31.1 & 51.6$\times$48.7 & 93.0 & 2.21 & 5.24 & 5.28 & 4.08 & 1.75 & H23, F18 \\
S15 & 4:14:12.33 & +28:08:46.6 & 36.0$\times$18.4 & 7.0 & 2.24 & 2.78 & 2.06 & 1.17 & 0.27 & H9, F6 \\
S16 & 4:18:33.13 & +28:27:31.0 & 85.8$\times$24.0 & 172.0 & 1.03 & 3.17 & 3.34 & 2.34 & 0.91 & None \\
S17 & 4:16:57.34 & +28:29:35.6 & 54.6$\times$32.0 & 38.0 & 1.44 & 2.96 & 2.61 & 1.84 & 1.04 & H58, F22 \\
S18 & 4:17:42.09 & +27:39:15.0 & 70.8$\times$26.0 & 75.0 & 2.70 & 4.94 & 4.01 & 2.97 & 0.86 & H13, F13 \\
S19 & 4:18:00.55 & +28:11:08.7 & 45.0$\times$22.2 & 165.0 & 0.55 & 1.70 & 1.74 & 1.23 & 0.52 & H45, F43 \\
S20 & 4:17:52.87 & +28:23:40.0 & 44.4$\times$37.1 & 160.0 & 1.16 & 2.08 & 1.78 & 1.28 & 0.89 & H18, F36 \\
S21 & 4:17:37.42 & +28:12:06.0 & 51.0$\times$35.4 & 27.0 & 1.40 & 3.06 & 2.61 & 1.67 & 0.73 & H44, F21 \\
S22 & 4:17:50.02 & +27:37:56.1 & 38.4$\times$28.8 & 140.0 & 0.58 & 1.31 & 1.09 & 0.81 & 0.36 & H34, F46 \\
S23 & 4:18:52.50 & +28:20:23.0 & 44.4$\times$24.8 & 99.0 & 5.47 & 4.95 & 2.75 & 1.53 & 0.33 & H3, F3 \\
S24 & 4:18:07.41 & +28:24:48.7 & 70.8$\times$30.9 & 30.0 & 1.92 & 3.87 & 3.15 & 1.94 & 0.64 & H10, F20 \\
S25 & 4:17:20.14 & +28:18:56.5 & 52.8$\times$40.1 & 12.0 & 2.76 & 3.84 & 2.51 & 1.42 & 0.71 & H16, F27 \\
\hline
\end{tabular}
\end{table*}

Three sets of sources were identified: sources in the 850-\um\ SCUBA-2 data (hereafter referred to as SCUBA-2 sources), sources in the 250-\um\ \emph{Herschel}-SPIRE data (hereafter referred to as \emph{Herschel} sources) and sources in the spatially-filtered 250-\um\ data (hereafter referred to as filtered-\emph{Herschel} sources).  Sources were identified using the source-finding algorithm CSAR \citep{kirk2013}.  CSAR is a dendrogram-based source-finding algorithm, which was run in its non-hierarchical mode on each of the three data sets. The criteria chosen for a robustly-detected source were a peak flux density $F_{\nu}^{peak}\geq 4\sigma$ and a minimum of a $3\sigma$ drop in flux density between adjacent sources, where $\sigma$ is the RMS noise level of the data (see \citealt{kirk2013}).  These stringent criteria were chosen in order to ensure that the sources we identified could be well-characterised, so that accurate comparisons could be made between cores detected in the different data sets.

We measured the $1\sigma$ RMS noise on the low-variance regions of the SCUBA-2 850-\um\ map to be 0.9$\,\pm\,$0.2 mJy/6-arcsec pixel and on the unfiltered 250-\um\ map to be 1.0$\,\pm\,$0.2 mJy/6-arcsec pixel.  Measuring the noise on the filtered 250-\um\ map produced values in the range 1.1--1.6\,mJy/6-arcsec pixel.  We adopted a value of 1.4\,$\pm$\,0.2\,mJy/6-arcsec pixel, as being representative.  The increase in RMS noise level in the filtered 250-\um\ map over the unfiltered 250-\um\ map is likely to be a result of the additional processing performed on the filtered map.

The regions of the SCUBA-2 map upon which source extraction was performed were those where the variance, as measured in the variance array, was $\leq 2$\,(mJy/6-arcsec pixel)$^{2}$.  These are the two large regions marked on Figures~\ref{fig:finding_chart1} and \ref{fig:finding_chart2}.  Noise levels across the \emph{Herschel} maps are more uniform than those across the SCUBA-2 map. However, all sources detected in the \emph{Herschel} data which were not fully located within one of the low-variance SCUBA-2 regions were excluded from further analysis, in order to allow an accurate comparison of the sources detected by the two instruments. 

We identified 26 sources in the 850-\um\ map, 211 sources in the 250-\um\ map and 140 sources in the filtered 250-\um\ map.  Sources smaller than the beam were rejected in the source extraction process. We examined the source samples to determine whether any sources contained protostars, based on whether they contained point sources at 70~\um\, or shorter wavelengths. One protostellar source was identified in the SCUBA-2 sample, along with 3 in the 250-\um\ sample and 2 in the filtered 250-\um\ sample.  These sources were excluded from further analysis, leaving us with 25 sources in the 850-\um\ map (shown on Figures~\ref{fig:finding_chart1} and \ref{fig:finding_chart2}), 208 sources in the unfiltered 250-\um\ map (shown on Figures~\ref{fig:finding_chart_A1} and \ref{fig:finding_chart_A2} in the online appendix), and 138 sources in the filtered 250-\um\ map (shown on Figures~\ref{fig:finding_chart_B1} and \ref{fig:finding_chart_B2} in the online appendix). The 25 SCUBA-2 sources (named S1--S25) are listed in Table~\ref{tab:scuba2_sources}, while the \emph{Herschel} sources (H1--H208) and filtered \emph{Herschel} sources (F1--F138) are presented along with the reduced SCUBA-2 data at the DOI listed in Section 2, above.  For each SCUBA-2 source, Table~\ref{tab:scuba2_sources} lists the name, right ascension and declination, measured major and minor FWHM sizes, position angle, flux densities as measured in filtered \emph{Herschel} 160\um, 250\um, 350\um\ and 500\um\ and SCUBA-2 850\um\ emission, and the equivalent sources in the \emph{Herschel} and filtered-\emph{Herschel} catalogues.  The details of the \emph{Herschel} tables are given at the same DOI.

\begin{figure*} 
\centering
\includegraphics[width=\textwidth]{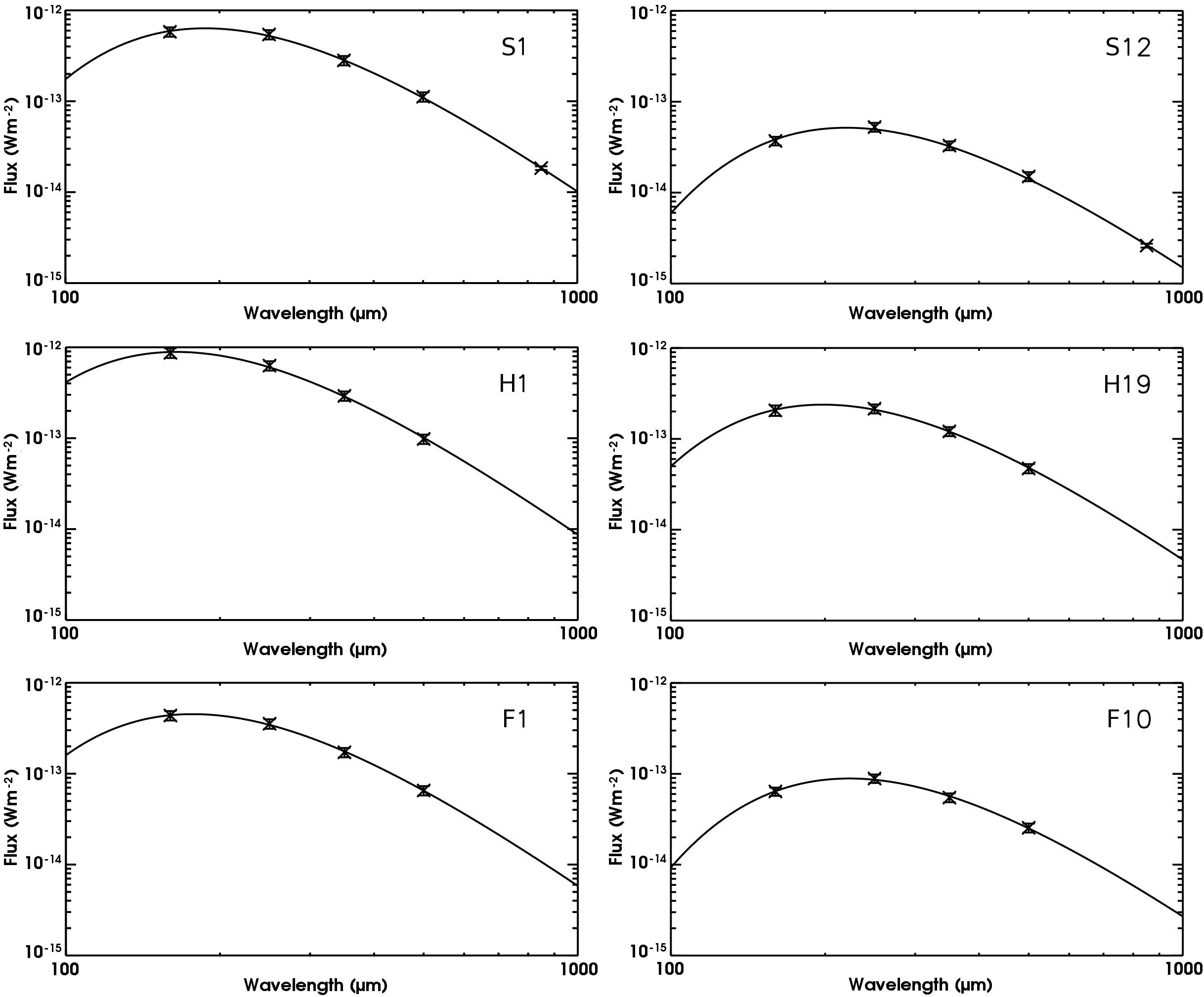}
\caption{SED fits for sources S1 and S12, and their counterparts in the \emph{Herschel} and filtered-\emph{Herschel} catalogues (H1 and F1, and H19 and F10, respectively).  See text for details.}
\label{fig:SEDs}
\end{figure*}

We derived temperatures and masses for each of our sources using the spectral energy distribution (SED) measured from the \emph{Herschel} and, in the case of the SCUBA-2 sources, the SCUBA-2 continuum data.  The flux densities of the SCUBA-2 sources were measured across 5 wavebands (filtered 160-\um, filtered 250-\um, filtered 350-\um, filtered 500-\um\ and 850-\um), all convolved to the SPIRE 500-\um\ resolution of 36 arcsec.  The flux densities of the \emph{Herschel} sources were measured across the \emph{unfiltered} 160-\um, 250-\um, 350-\um\ and 500-\um\ wavebands, while the flux densities of the filtered-\emph{Herschel} sources were measured across the \emph{filtered} 160-\um, 250-\um, 350-\um\ and 500-\um\ wavebands, in both cases also at 36-arcsec resolution.  We convolved the maps to a common 36-arcsec resolution using the \emph{Herschel}$\to$\emph{Herschel} and SCUBA-2$\to$\emph{Herschel} convolution kernels described by \citet{pattle2015}.  Flux densities for each source were measured using elliptical apertures with major and minor axes of twice the major and minor FWHM returned by CSAR.  This would enclose 99.5\% of the flux in a Gaussian distribution.

Note that the 850-\um\ flux densities listed in Table~\ref{tab:scuba2_sources} do not have the SCUBA-2 aperture photometry corrections discussed by \citet{dempsey2013} applied to them.  The SCUBA-2 aperture photometry corrections are determined for point sources, and account for flux in the secondary beam of the JCMT not enclosed by a small aperture (the JCMT's secondary beam has a FWHM of 48\,arcsec at 850~\um; see \citealt{dempsey2013}).  We do not use these aperture photometry corrections in this work, as their applicability to either extended sources or non-circular apertures is not certain.  As noted above, the major and minor axes of the apertures which we use are twice the FWHM values of the source listed in Table~\ref{tab:scuba2_sources}, and hence the smallest of our sources have their fluxes measured in apertures with minor axes $\sim 40$\,arcsec.  Hence, for the smallest of our sources, the 850-\um\ flux density may be underestimated by up to a maximum of 10\,per\,cent.  However, for the large majority of our sources, the aperture photometry correction should be $\lesssim 3$\,per\,cent.  We direct the reader to \citet{dempsey2013} for further information.

\begin{table}
\centering
\caption{Derived properties of the sources found by SCUBA-2 in the L1495 region -- temperature, mass, density and mean deconvolved full-width at half-maximum (geometric mean of major and minor axes with beam-size subtracted in quadrature).}
\label{tab:properties}
\begin{tabular}{c c c c c}
\hline
Source & & & H$_{2}$ & Deconv. \\
Index & Temp. & Mass & Density & FWHM \\
 & (K) & (M$_{\odot}$) & ($\times10^{4}$\,cm$^{-3}$) & (pc) \\
\hline
S1 & 14.9$\pm$0.4 & 0.613$\pm$0.267 & 6.34$\pm$2.76 & 0.032 \\
S2 & 10.1$\pm$0.2 & 0.256$\pm$0.057 & 6.94$\pm$1.56 & 0.023 \\
S3 & 10.9$\pm$0.2 & 0.420$\pm$0.114 & 3.96$\pm$1.08 & 0.033 \\
S4 & 11.2$\pm$0.2 & 0.156$\pm$0.034 & 10.55$\pm$2.29 & 0.017 \\
S5 & 10.1$\pm$0.2 & 0.336$\pm$0.081 & 8.08$\pm$1.94 & 0.024 \\
S6 & 11.2$\pm$0.2 & 0.120$\pm$0.025 & 7.04$\pm$1.48 & 0.018 \\
S7 & 10.6$\pm$0.2 & 0.110$\pm$0.023 & 6.88$\pm$1.41 & 0.018 \\
S8 & 12.0$\pm$0.3 & 0.391$\pm$0.116 & 3.37$\pm$1.00 & 0.034 \\
S9 & 11.1$\pm$0.2 & 0.324$\pm$0.082 & 3.95$\pm$1.00 & 0.030 \\
S10 & 14.4$\pm$0.4 & 0.126$\pm$0.029 & 2.47$\pm$0.58 & 0.026 \\
S11 & 9.3$\pm$0.2 & 0.196$\pm$0.043 & 9.38$\pm$2.04 & 0.019 \\
S12 & 12.7$\pm$0.3 & 0.113$\pm$0.024 & 3.27$\pm$0.70 & 0.023 \\
S13 & 14.3$\pm$0.4 & 0.079$\pm$0.017 & 2.94$\pm$0.64 & 0.021 \\
S14 & 10.9$\pm$0.2 & 0.355$\pm$0.092 & 3.05$\pm$0.79 & 0.034 \\
S15 & 16.1$\pm$0.5 & 0.026$\pm$0.005 & 1.67$\pm$0.34 & 0.017 \\
S16 & 10.9$\pm$0.2 & 0.182$\pm$0.039 & 2.11$\pm$0.45 & 0.031 \\
S17 & 11.0$\pm$0.2 & 0.205$\pm$0.047 & 3.03$\pm$0.70 & 0.028 \\
S18 & 12.9$\pm$0.3 & 0.124$\pm$0.027 & 1.69$\pm$0.37 & 0.029 \\
S19 & 10.7$\pm$0.2 & 0.109$\pm$0.022 & 3.72$\pm$0.76 & 0.021 \\
S20 & 10.7$\pm$0.2 & 0.186$\pm$0.042 & 3.02$\pm$0.69 & 0.028 \\
S21 & 11.9$\pm$0.2 & 0.122$\pm$0.026 & 1.73$\pm$0.37 & 0.029 \\
S22 & 11.4$\pm$0.2 & 0.065$\pm$0.013 & 1.90$\pm$0.38 & 0.023 \\
S23 & 19.6$\pm$0.7 & 0.023$\pm$0.005 & 0.69$\pm$0.15 & 0.023 \\
S24 & 12.9$\pm$0.3 & 0.090$\pm$0.018 & 0.95$\pm$0.20 & 0.032 \\
S25 & 13.4$\pm$0.3 & 0.095$\pm$0.021 & 1.05$\pm$0.23 & 0.031 \\
\hline
\end{tabular}
\end{table}

The SED of each source was fitted with a modified blackbody distribution (see Figure~\ref{fig:SEDs}), in order to determine the mean, column-density-weighted, line-of-sight dust temperature.  The monochromatic flux density $F_{\nu}$ is given at frequency $\nu$ by
\begin{equation}
\lambda F_{\lambda} = \nu F_{\nu} = \nu\Omega f B_{\nu}(T) \left(1 -
e^{-\left(\frac{\nu}{\nu_{c}}\right)^{\beta}}\right) \, ,
\label{eq:sed}
\end{equation}
\noindent
where $B_{\nu}(T)$ is the Planck function at dust temperature $T$, $\Omega$ is the solid angle of the aperture, $f$ is the filling factor of the source in the aperture, and $\beta$ is the dust emissivity index. The frequency at which the optical depth becomes unity is taken to be the canonical value, $\nu_{c}=6\,$ THz \citep{wardthompson2002a}, as we do not have enough data points to accurately constrain a fourth parameter in our fitting process.

We determined a typical dust emissivity index ($\beta$), which we adopted for our cores in order to more accurately constrain the SED fitting process.  While temperature and dust emissivity index can be fitted simultaneously using SCUBA-2 850\um\ data in conjunction with the \emph{Herschel} photometric bands (see \citealt{sadavoy2013}), the \emph{Herschel} data alone, covering the wavelength range 160 -- 500~\um, do not provide the long-wavelength information necessary to accurately constrain both parameters.  We decided to use a fixed value of $\beta$ when deriving best-fit temperatures for all of our cores, including those with an 850-\um\ detection, in order to make a fair comparison between the different sets of cores.  In order to find a suitable $\beta$, we fitted SEDs to the subset of the filtered-\emph{Herschel} cores with detections in 850-\um\ emission. We determined an SED using only the filtered \emph{Herschel} fluxes for each source, and from the best-fit SED predicted an 850-\um\ flux density.  These predicted 850-\um\ flux densities were then compared to the values measured in the SCUBA-2 850-\um\ map.  We repeated this process for a range of values of $\beta$, as well as allowing $\beta$ to vary as a free parameter, in order to determine the value that best predicted the 850-\um\ flux densities.

For each of the values of $\beta$ tested, we calculated the mean reduced $\chi^{2}$ value of the SED fits.  We also calculated the reduced $\chi^{2}$ value of the 1:1 relation between predicted and measured 850-\um\ flux densities. We found that when emissivity index was allowed to vary as a free parameter, the 850-\um\ flux densities were generally well-predicted.  We found the mean value of $\beta$ when it was varied as a free parameter to be 1.3, and the standard deviation on this value to be 0.6.  When $\beta$ was fixed, we found that values in the range $\beta=$1.1--1.4 gave indistinguishably good results.  We therefore chose to adopt a value of dust emissivity index $\beta=1.3\pm0.6$ for the remainder of this work.

A dust emissivity index of 1.3 is lower than that typically expected for a starless core. A wide variety of $\beta$ values have been determined for starless cores.  For example, \citet{shirley2005} found $\beta=$1.8--1.9 for the starless core L1498; \citet{friesen2005} found $\beta=$1.3--2.1 for a sample of hot starless cores; \citet{schnee2010} found $\beta=2.2\pm0.6$ for the starless core TMC-1c; and \citet{sadavoy2013} found $\beta=$1.6--2.0 toward cores in the Perseus molecular cloud. Our low value of $\beta$ may be due to there being multiple dust temperature components along the lines of sight toward our sources, which would broaden the SEDs of the sources, and hence lower their apparent $\beta$ values.  This effect was discussed in detail by \citet{martin2012}.  Starless cores are not expected to be isothermal sources: a temperature gradient from $~15\,$K at their edges to $~7\,$K in their heavily-shielded centres has been seen elsewhere (e.g. \citealt{stamatellos2007}), even in the absence of external heating sources. The surrounding, more tenuous, material of the molecular cloud will be warmer still.  While the spatial filtering of the SCUBA-2 observations will to some extent ameliorate this effect, there will still be some material along the line of sight not associated with the core.

Our low $\beta$ might alternatively, or additionally, be the result of grain growth within the densest regions of the starless cores. Our value of $\beta$ is intermediate between the value expected in molecular clouds of $\beta=$1.5--2.0 (e.g. \citealt{draine2007}; \citealt{draine1984}), and $\beta=1.0$, expected in protoplanetary discs \citep{beckwith1991}.

We fitted our sources using $\beta=1.3$ in order to determine their dust temperatures.  We then determined our source masses using the \citet{hildebrand1983} relation
\begin{equation}
M = \frac{F_{\nu}(850\upmu{\rm m})D^{2}}{\kappa_{\nu(850\upmu{\rm m})}B_{\nu(850\upmu{\rm m})}(T)}\, ,
\label{eq:mass}
\end{equation}
\noindent
where $F_{\nu}(850\upmu{\rm m})$ is the flux density at 850~\um.  For the SCUBA-2 sources, $F_{\nu}(850\upmu{\rm m})$ was taken to be the measured SCUBA-2 850-\um\ flux density of the source, while for the \emph{Herschel} and filtered-\emph{Herschel} sources, $F_{\nu}(850\upmu{\rm m})$ was taken to be the flux density at 850\um\ as extrapolated from the best-fit SED. $D$ is the distance to Taurus (140\,pc), $B_{\nu(850\upmu{\rm m})}(T)$ is the Planck function, and $\kappa_{\nu(850\upmu{\rm m})}$ is the dust mass opacity, as parameterised by \citet{beckwith1990}: $\kappa_{\nu}\,=\,0.1(\nu/10^{12}{\rm Hz})^{\beta}$\,cm$^{2}$g$^{-1}$ (assuming a standard dust-to-gas mass ratio of 1:100).  Again, the dust emissivity index $\beta$ was taken to be 1.3.  Some example SED fits for sources with counterparts in all three catalogues are shown in Figure~\ref{fig:SEDs}.

We calculated the mean volume density of molecular hydrogen for each source as 
\begin{equation}
  n({\rm H}_{2})=\frac{M}{\mu m_{\textsc{h}}}\frac{1}{\frac{4}{3}\pi R^{3}},
  \label{eq:density}
\end{equation}
\noindent
where $R$ is the equivalent deconvolved FWHM of the source.  The equivalent FWHM was taken to be the geometric mean of the deconvolved major and minor FWHMs, as determined by CSAR.  The deconvolution assumed a beam size of 14.1 arcsec in the case of SCUBA-2 sources and 18.1 arcsec in the case of \emph{Herschel} and filtered-\emph{Herschel} sources. The mean molecular weight $\mu$ was taken to be 2.86, assuming that the gas is $\sim 70$\% H$_{2}$ by mass \citep{kirk2013}.

\section{Discussion}

\begin{sidewaysfigure*} 
\centering
\textcolor{white}{\rule{0.99\textwidth}{0.80\textwidth}}
\includegraphics[width=\textwidth]{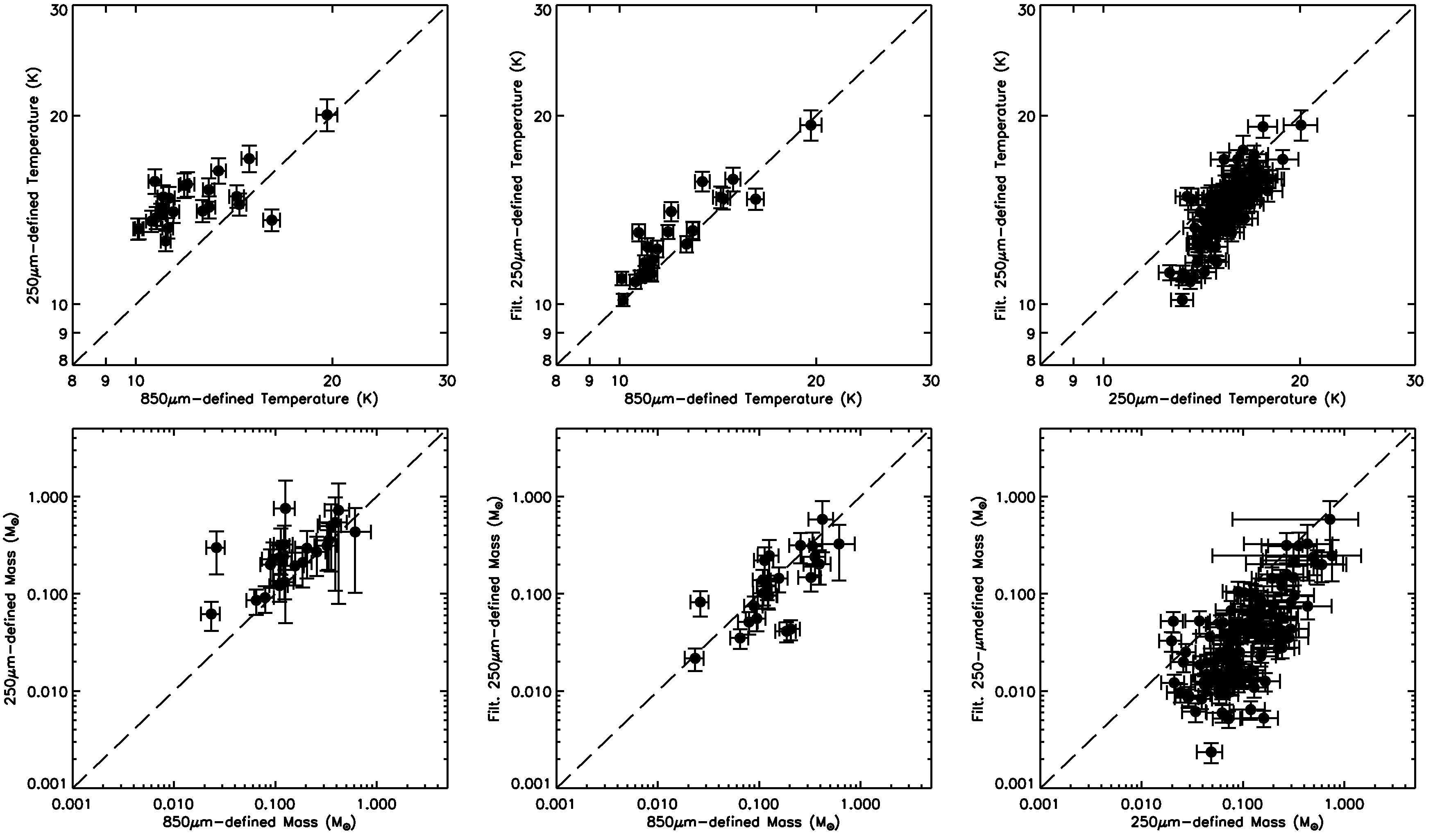}
\begin{minipage}{\textwidth}
\medskip
\end{minipage}
\begin{minipage}{\textwidth}
\textbf{Figure 6.} Comparison of the properties of 23 SCUBA-2 sources with their counterpart \emph{Herschel} sources (left-hand column), and filtered-\emph{Herschel} sources (middle column). The 118 sources in common between the \emph{Herschel} and filtered-\emph{Herschel} catalogues are compared in the right-hand column. The parameters compared are temperature (top row) and mass (second row).  For temperature and mass the filtered-\emph{Herschel} sources match the SCUBA-2 sources better than the unfiltered \emph{Herschel} sources, showing that the filtering process makes the \emph{Herschel} data more comparable to the SCUBA-2 data. The filtered data appear to underestimate, on average, the temperature and mass relative to the unfiltered data (see text for discussion).
\end{minipage}
\label{fig:comparison}
\end{sidewaysfigure*}

The temperatures, masses, densities and sizes of the cores detected by SCUBA-2 are listed in Table~\ref{tab:properties}. The same information for \emph{Herschel} and filtered-\emph{Herschel} sources is listed in Tables~A3 and~A4 in the online appendix. The properties of those sources with counterparts in another catalogue are compared with the properties of their counterparts in Figures~\ref{fig:comparison} and~\ref{fig:comparison_2}.   A source's counterpart in another catalogue is the nearest neighbour to the source in that catalogue, provided that the source and its neighbour are separated by less than the FWHM of the larger of the two sources.  It can be seen from the central column of plots in Figures~\ref{fig:comparison} and~\ref{fig:comparison_2} that sources found in both the SCUBA-2 and the filtered-\emph{Herschel} catalogues are typically found to have very similar properties. This tends to indicate that the filtering process is the correct method by which to make SCUBA-2 and \emph{Herschel} data comparable.

\addtocounter{figure}{1}

\begin{sidewaysfigure*} 
\centering
\textcolor{white}{\rule{0.99\textwidth}{0.80\textwidth}}
\includegraphics[width=\textwidth]{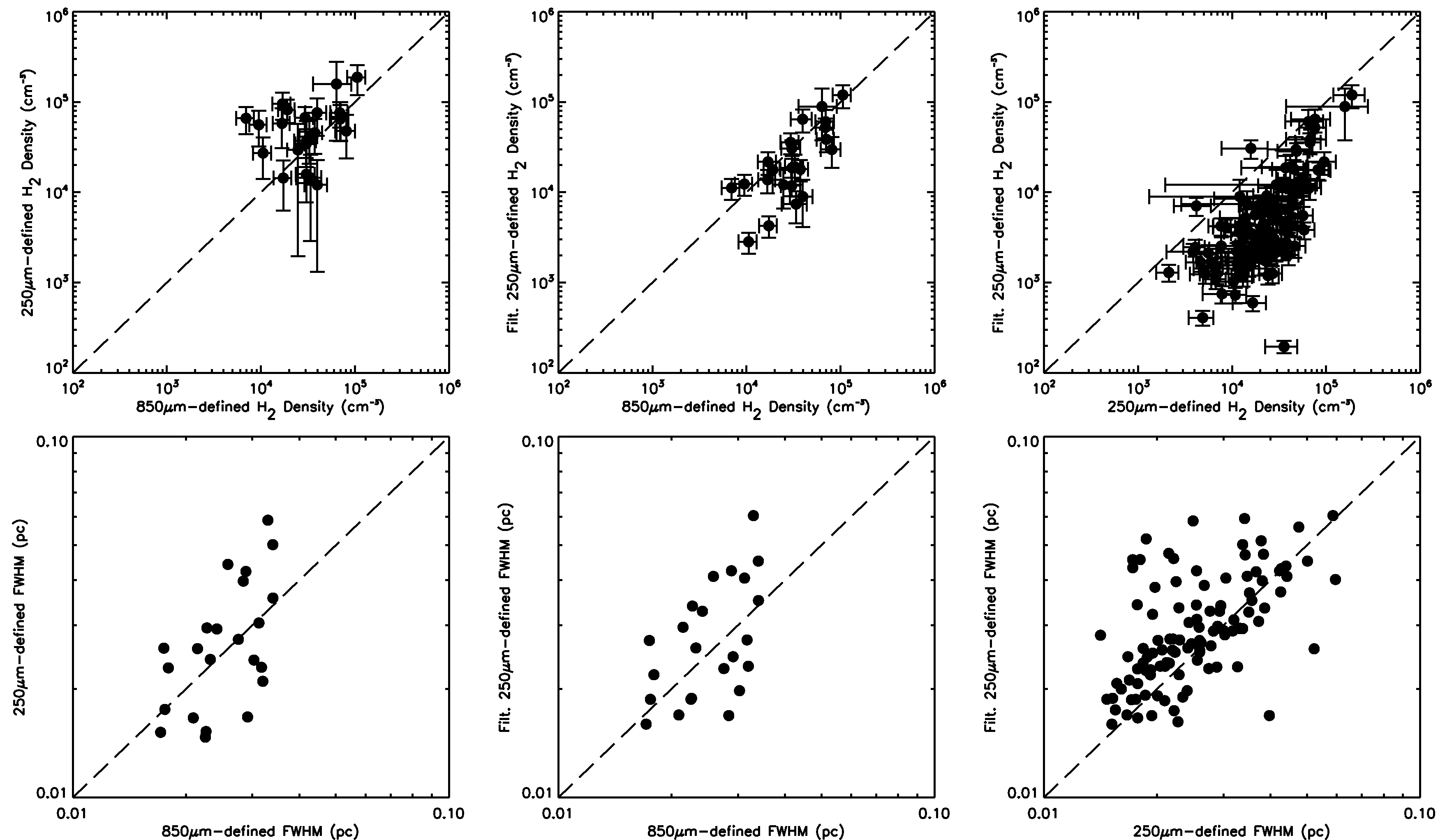}
\begin{minipage}{\textwidth}
\medskip
\end{minipage}
\begin{minipage}{\textwidth}
\textbf{Figure 7.} Comparison of the properties of 23 SCUBA-2 sources with their counterpart \emph{Herschel} sources (left-hand column), and filtered-\emph{Herschel} sources (middle column). The 118 sources in common between the \emph{Herschel} and filtered-\emph{Herschel} catalogues are compared in the right-hand column. The parameters compared are density (top row) and deconvolved FWHM size (second row). For density the filtered-\emph{Herschel} sources match the SCUBA-2 sources better than the unfiltered \emph{Herschel} sources. The filtered data appear to underestimate, on average, density relative to the unfiltered data (see text for discussion). Source FWHM size appears unaffected by the filtering process for typical core sizes.
\end{minipage}
\label{fig:comparison_2}
\end{sidewaysfigure*}
\addtocounter{figure}{1}

\begin{figure*} 
\centering
\includegraphics[width=0.75\textwidth]{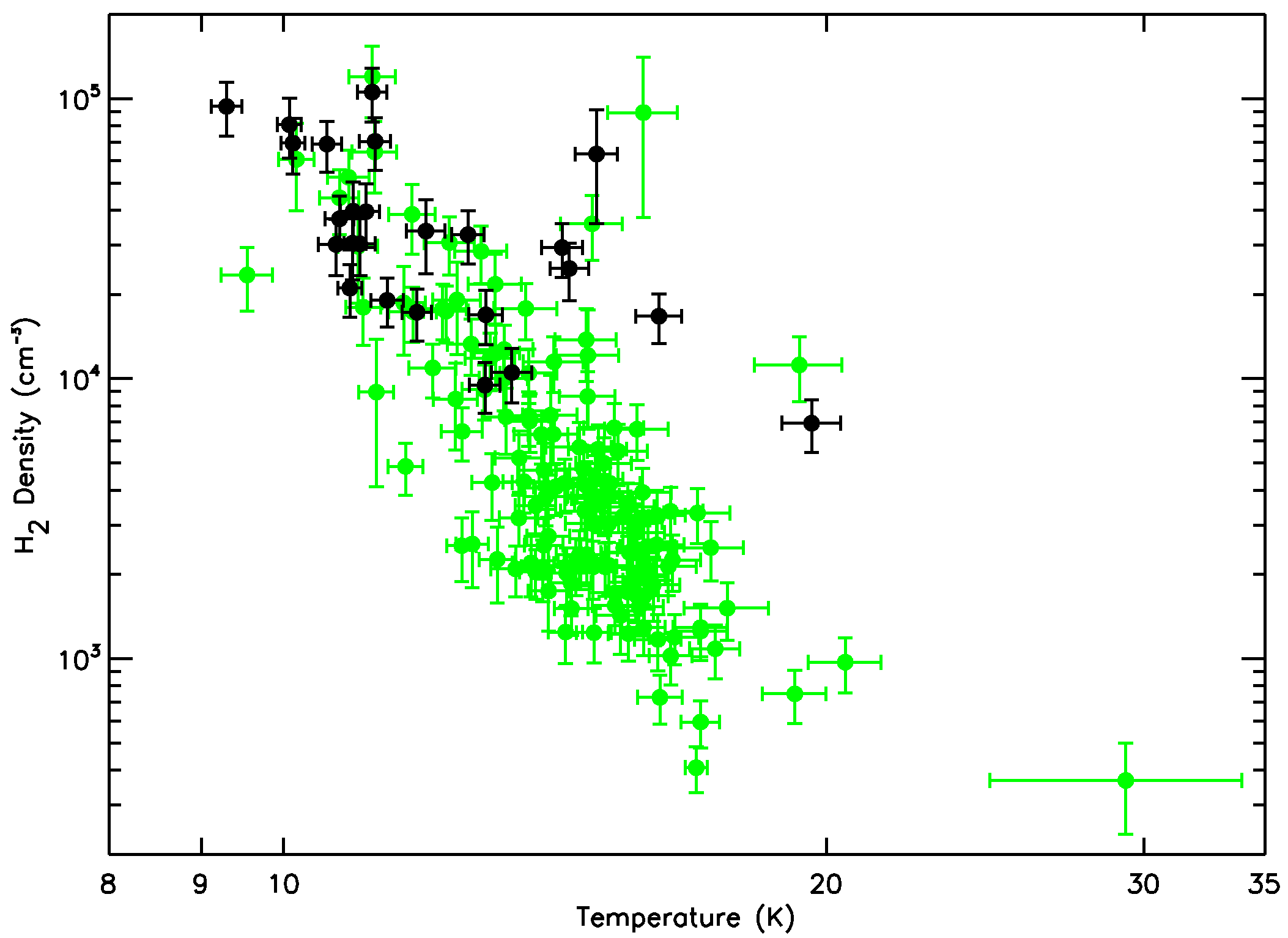}
\caption{Plot of core density against temperature.  Black symbols are SCUBA-2 sources, green symbols are filtered-\emph{Herschel} sources. There appears to be a cut-off in density, with the minimum SCUBA-2 source density being $\sim 6\times10^{3}$ H$_{2}$/cm$^{3}$, while the lowest-density filtered-\emph{Herschel} sources are $\sim 4\times10^{2}$ H$_{2}$/cm$^{3}$. However, the filtered-\emph{Herschel} sources and the SCUBA-2 sources have a similar range in temperature, $\sim 9$--20\,K, with no apparent cut-off.}
\label{fig:temp_density}
\end{figure*}

\emph{Herschel} sources are typically measured to be warmer than their SCUBA-2 and filtered-\emph{Herschel} counterparts. We hypothesise that this is due to the filtering process removing much of the extended foreground and background emission originating from warmer material, thus reducing the line-of-sight temperature determined by the fitting process.  Source FWHM size is typically not strongly affected by the filtering process, with no clear propensity for cores to become smaller or larger.  This suggests that CSAR is identifying the core material accurately: in the unfiltered case CSAR is identifying emission peaks above the background large-scale emission of the cloud, while in the filtered case, the background large-scale emission has been removed, and CSAR is identifying emission peaks above the noise in the data.

\emph{Herschel} sources are typically more massive than their counterpart filtered-\emph{Herschel} sources.  Again, this is due to the removal of large-scale signal by the filtering process.  Without the additional contribution of the background signal, the total flux measured in the filtered-\emph{Herschel} apertures will be less than that measured in corresponding \emph{Herschel} apertures (typically similarly-sized, as discussed above).

\emph{Herschel} sources are typically denser than their counterpart filtered-\emph{Herschel} sources.  This is a counterintuitive result, as the removal of low-density large-scale structure might be expected to leave behind only the higher-density cores.  However, as discussed above, the size of the cores detected by CSAR is largely unaffected by the filtering process, which is likely due to CSAR detecting the core itself, rather than the material in which it is embedded.  Hence, a given core detected in both \emph{Herschel} and filtered-\emph{Herschel} emission will be the same size in both cases, but will be less massive, and so less dense, in the filtered-\emph{Herschel} case.

These results suggest that the measures of core temperature, mass and density determined from the filtered-\emph{Herschel} data are more representative of the true properties of the starless cores than those determined from the unfiltered \emph{Herschel} data, as in the filtered-\emph{Herschel} case the flux measured should be that of the core itself, rather than being the flux of both the core and the full line-of-sight column of material in which it is embedded.

Figure~\ref{fig:temp_density} shows the relationship between temperature and density for the filtered-\emph{Herschel} sources and the SCUBA-2 sources.  The SCUBA-2 sources follow the same density-temperature relation as the filtered-\emph{Herschel} sources, $n({\rm H}_{2})\propto T^{-8.5\pm 0.5}$. However, only the densest filtered-\emph{Herschel} sources have a counterpart SCUBA-2 source, with the minimum SCUBA-2 source density being $\sim 3\times 10^{-17}$\,kg\,m$^{-3}$ ($\sim 6\times10^{3}$ H$_{2}$/cm$^{3}$), while the lowest-density filtered-\emph{Herschel} sources are $\sim 2\times 10^{-18}$\,kg\,m$^{-3}$ ($\sim 4\times10^{2}$ H$_{2}$/cm$^{3}$). The filtered-\emph{Herschel} sources and the SCUBA-2 sources have a similar range of temperatures, being $\sim 9$--20\,K.

Those SCUBA-2 sources which do not follow the temperature-density relation are S1 (L1495A-S), S10, S13, S15 and S23.  All of these sources are significantly ($\sim 5$\,K) warmer than might be expected from their density.  Of these five sources, S1 and S23 are heated by V892 Tau (IRAS04155+2812), as discussed above.  Sources S10 and S13 are being heated by IRAS04113+2758, while source S15 is associated with the IR source IRAS04111+2800G, and may in fact have an embedded YSO within it, possibly a VeLLO (Very Low Luminosity Object -- \citealt{young2004}).

\begin{figure*} 
\centering
\includegraphics[width=0.75\textwidth]{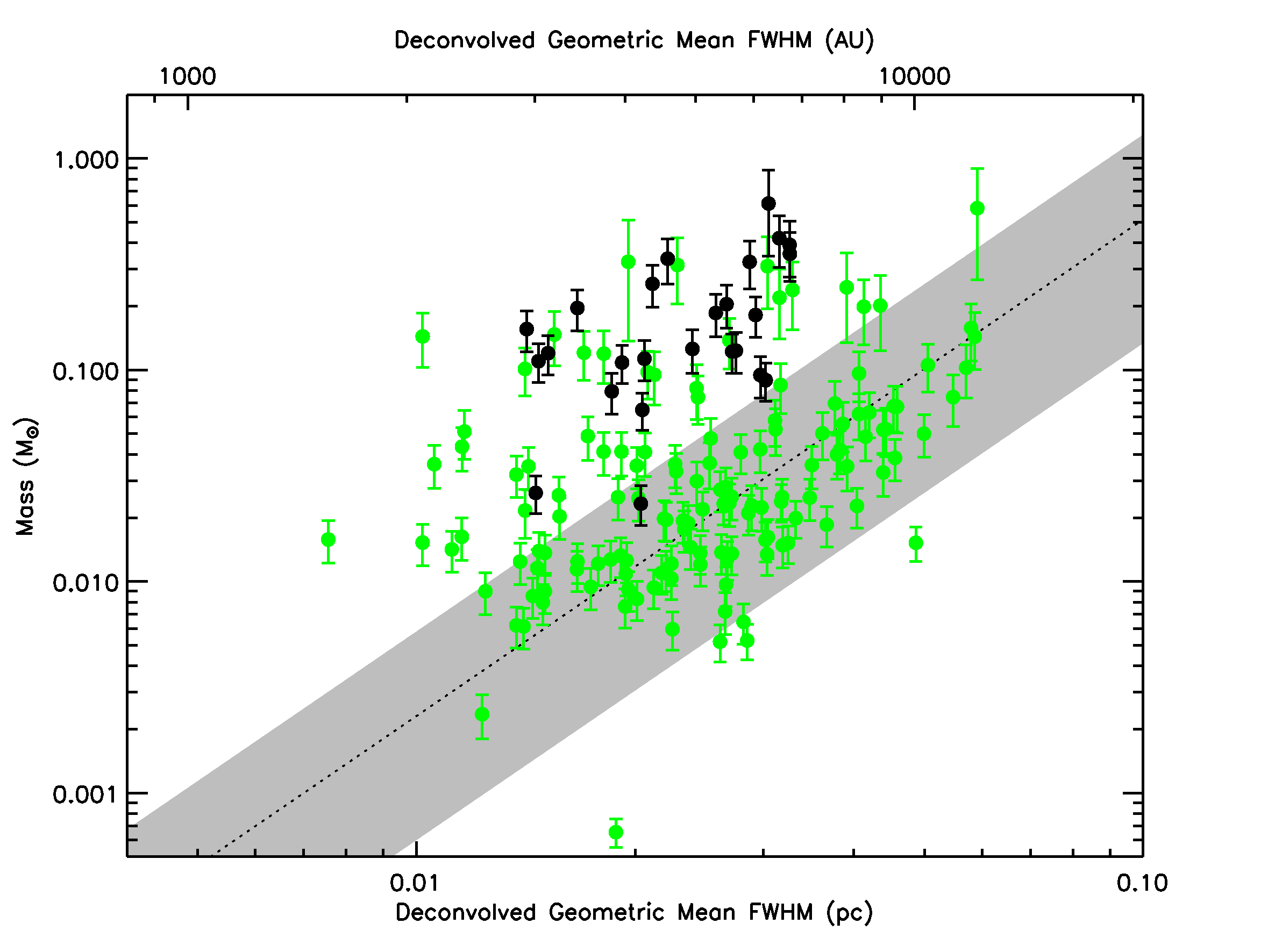}
\caption{Plot of core mass against deconvolved size.  Black symbols are SCUBA-2 sources, green symbols are filtered-\emph{Herschel} sources. There appears to be no tendency for the SCUBA-2 sources to be smaller in size than the filtered-\emph{Herschel} sources. However, the SCUBA-2 sources are among the most massive.  The grey band indicates the region which unbound starless cores are expected to inhabit (\citealt{elmegreen1996}; \citealt{andre2010}).  A substantial fraction of the filtered-\emph{Herschel} sources lie within this region.  However, the SCUBA-2 sources in almost all cases occupy the region above this, in which prestellar cores are expected to be found (c.f. \citealt{andre2010}).}
\label{fig:radius_mass}
\includegraphics[width=0.75\textwidth]{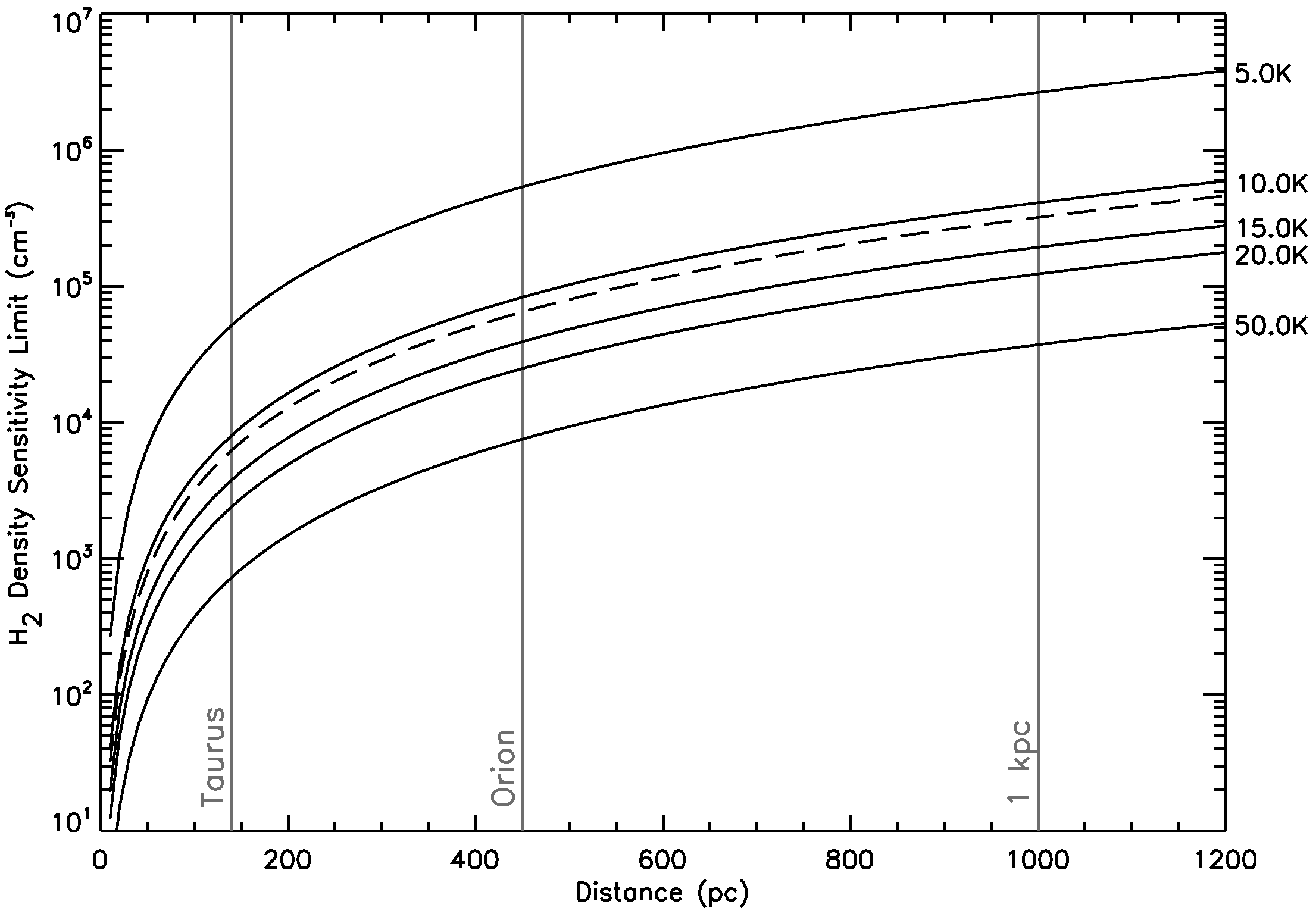}
\caption{Plot of the lowest source density detectable using SCUBA-2 Internal Release 1 GBS data, as a function of distance, for various assumed source temperatures.  The dashed line shows the SCUBA-2 GBS density sensitivity limit as a function of distance at the mean temperature of the non-externally heated starless cores in our sample (11.3\,K). All of the relations are normalised against a density limit of $\sim 6.3\times10^{3}$ H$_{2}$/cm$^{3}$ at a temperature of 11.3\,K and a distance of 140\,pc, as we find for the mean of the cores in Taurus in this work.}
\label{fig:density_sensitivity}
\end{figure*}

Figure~\ref{fig:radius_mass} shows the mass-size relation for the filtered-\emph{Herschel} and SCUBA-2 sources.  There is no tendency for the SCUBA-2 sources to be smaller in size than the filtered-\emph{Herschel} sources. However, the SCUBA-2 sources are among the most massive.  The grey band shown on Figure~\ref{fig:radius_mass} indicates the region of the mass/size plane which unbound, transient starless cores are expected to inhabit (\citealt{elmegreen1996}; \citealt{andre2010}).  A substantial fraction of the filtered-\emph{Herschel} sources lie within this region.  However, the SCUBA-2 sources in almost all cases occupy the region of the mass/size plot in which gravitationally bound prestellar cores are expected to be found (c.f. \citealt{andre2010}).

Figure~\ref{fig:temp_density} shows a clear cutoff in density below which no SCUBA-2 sources are detected. However, there appears to no similar cutoff in temperature.  Figure~\ref{fig:radius_mass} shows no tendency for SCUBA-2 sources to be smaller in radius than filtered-\emph{Herschel} sources.  This indicates that the criterion for determining whether a source detected using \emph{Herschel} will be detectable with SCUBA-2 is its mass for a given size, or in other words, density.

This corresponds, in terms of the measurable parameters, effectively to surface brightness. For a given temperature, the higher column density material will produce a higher surface brightness. Also, for non-elongated geometries, higher column density corresponds to higher volume density. Therefore, it appears that it is not mass or size alone that determines detectability with SCUBA-2, but rather a combination of mass and size that corresponds to density.  Dust at temperatures typically found in starless cores emits more flux at \emph{Herschel} wavelengths than at SCUBA-2 850-\um\ wavelength, because \emph{Herschel} wavelengths coincide with the peak of the black-body function, while 850\um\ is on the Rayleigh-Jeans tail. Hence, for roughly the same absolute noise level (which these maps have), any given core will appear at a higher signal-to-noise ratio to \emph{Herschel} than to SCUBA-2. The exceptions to this are those sources with associated or nearby stars or protostars, because in those cases the above assumption, that everything is at roughly similar temperature, breaks down.

We test this hypothesis further in Appendix B, by scaling the RMS noise on the filtered-\emph{Herschel} data such that the relative, rather than the absolute, noise levels of the SCUBA-2 850-\um\ and filtered-\emph{Herschel} 250-\um\ data are comparable.  We show in Appendix B that the sources detected in the increased-noise filtered-\emph{Herschel} data are preferentially the denser sources, as expected.

Figure~\ref{fig:density_sensitivity} shows how the minimum density $n$ to which SCUBA-2 is sensitive varies as a function of distance and temperature.  A core of temperature $T$ at distance $D$ will be detectable in SCUBA-2 GBS data if its density is greater than or equal to $n$, where
\begin{equation}
  n=n_{0}\left(\frac{D}{D_{0}}\right)^{2}\frac{e^{\nicefrac{h\nu}{k_{{\mbox{\scriptsize{\textsc{b}}}}}T}}-1}{e^{\nicefrac{h\nu}{k_{{\mbox{\scriptsize{\textsc{b}}}}}T_{0}}}-1}
  \label{eq:density_sensitivity}
\end{equation}
This relation is normalised to a density sensitivity $n_{0}=6.3\times10^{3}$ H$_{2}$/cm$^{3}$ at the canonical distance of the Taurus molecular cloud of $D_{0}=140$\,pc and the mean temperature of our non-externally-heated starless cores, $T_{0}=11.3$\,K.  The density sensitivity limit at a given distance decreases as source temperature increases.

The distances to the Taurus and Orion molecular clouds are marked on Figure~\ref{fig:density_sensitivity}.  The density sensitivity limits for a 10\,K core at 140\,pc (close to the typical core temperature in Taurus) and a 50\,K core at 450\,pc (a typical core temperature in the high-mass star-forming region Orion, shown for comparison) are very similar: $8\times10^{3}$ H$_{2}$/cm$^{3}$ and $7\times10^{3}$ H$_{2}$/cm$^{3}$ respectively.  Hence, their relative detectability in SCUBA-2 GBS data should be similar.

It is important to note that the SCUBA-2 sensitivity limits found in this work are those of the JCMT GBS Internal Release 1 data reduction, and not an intrinsic property of data from the instrument.  It is possible for further data to be taken, improving the sensitivity, until the confusion limit is reached.  Similarly, the \emph{Herschel} data are at the (fixed) sensitivity of the HGBS, and not necessarily confusion-limited.  Furthermore, as discussed above, we employed a stringent set of source selection criteria when creating our catalogue.  However, this does not alter our conclusion that, for SCUBA-2 and \emph{Herschel} maps treated identically and with sources extracted in the same manner, the surface brightness effect from which the density sensitivity threshold results will result in SCUBA-2 picking out the densest subset of the cores detected by \emph{Herschel}, and hence those most likely to be pre-stellar in nature.

\section{Conclusions}

In this paper we have compared data of the Taurus L1495 region from JCMT-SCUBA-2 at 850~\um\ with \emph{Herschel} data from 160 to 500\um. We have spatially filtered the \emph{Herschel} data to match the data processing carried out on the SCUBA-2 data.

We have extracted, and characterised the properties of, starless and pre-stellar cores from the SCUBA-2 850\um, \emph{Herschel}-SPIRE 250\um, and spatially-filtered \emph{Herschel} 250\um\ data, and have compared the cores found in the different data sets. Our goal was to determine which property of a starless core identified by \emph{Herschel} is most important in determining whether the same core would be detected with SCUBA-2.

We identified sources using the CSAR extraction algorithm.  We detected 25 sources in the SCUBA-2 850-\um\ data, 208 sources in the \emph{Herschel} 250-\um\ data, and 138 sources in the spatially-filtered \emph{Herschel} 250\um\ data.

We determined a representative dust emissivity index of our sources of $\beta=1.3\pm0.6$.  This was the value of $\beta$ which best predicted the SCUBA-2 850-\um\ flux densities of our sources from their spectral energy distribution (SED) in filtered \emph{Herschel} emission. We determined mean line-of-sight temperatures for our sources by SED fitting. This then allowed an accurate mass determination to be made for each source.

We found that cores detected by SCUBA-2 and cores detected in filtered 250-\um\ emission have similar properties, obeying the same temperature-density relation.  Cores extracted from, and characterised using, unfiltered \emph{Herschel} data typically have higher temperatures and densities than their counterparts in the SCUBA-2 data.  This is due to extended emission along the line of sight, which is removed by the filtering process.  This further confirmed that spatial filtering was necessary to accurately compare SCUBA-2 and \emph{Herschel} data.

We found that SCUBA-2 detects only the densest starless cores, with no SCUBA-2 cores having densities below $\sim 6 \times 10^{3}$ H$_{2}$/cm$^{3}$, an order of magnitude higher density than the least dense filtered 250-\um-detected \emph{Herschel} core.  There is no equivalent cutoff in temperature, with both SCUBA-2 and \emph{Herschel} sources having temperatures in the range $\sim 9$--20\,K.  Neither are SCUBA-2 cores typically smaller in radius than \emph{Herschel} cores -- i.e. the spatial filtering introduced by SCUBA-2 does not appear to change the measured FWHM of a starless core observed at this distance.

Thus, we found that the criterion for whether a starless or prestellar core detected in \emph{Herschel} data will also be detected in SCUBA-2 data is its density (for a given temperature). In the case of Taurus, for SCUBA-2 GBS Internal Release 1 data, this was $\sim 6\times 10^{3}$ H$_{2}$/cm$^{3}$. This corresponds to a cut-off in surface brightness, below which SCUBA-2 is no longer sensitive. This makes SCUBA-2 ideal for selecting those cores in \emph{Herschel} catalogues that are closest to forming stars. This information can be used in the analysis of SCUBA-2 and \emph{Herschel} data of other GBS regions, as shown in Figure~\ref{fig:density_sensitivity}, and to plan future SCUBA-2 observing campaigns.

\section*{Acknowledgements}

The JCMT has historically been operated by the Joint Astronomy Centre on behalf of the Science and Technology Facilities Council of the United Kingdom, the National Research Council of Canada, and the Netherlands Organisation for Scientific Research.  Additional funds for the construction of SCUBA-2 were provided by the Canada Foundation for Innovation.  \emph{Herschel} is an ESA space observatory with science instruments provided by European-led Principal Investigator consortia and with important participation from NASA.  DWT wishes to thank STFC for FEC support under grant number ST/K002023/1.  KP wishes to thank STFC for postdoctoral support under grant numbers ST/K002023/1 and ST/M000877/1 and studentship support under grant number ST/K501943/1.

\bibliographystyle{mn2e_fix}

\vspace{1cm}

\textit{
$^1$Jeremiah Horrocks Institute, University of Central Lancashire, Preston PR1 2HE, UK \\
$^2$School of Physics \& Astronomy, Cardiff University, The Parade, Cardiff CF24 3AA, UK \\
$^3$Astrophysics Group, Cavendish Laboratory, J J Thomson Avenue, Cambridge CB3 0HE, UK \\
$^4$Kavli Institute for Cosmology, Institute of Astronomy, University of Cambridge, Madingley Road, Cambridge CB3 0HA, UK \\
$^5$Department of Physics and Astronomy, University of Exeter, Stocker Road, Exeter EX4 4QL, UK \\
$^6$NRC Herzberg Astronomy and Astrophysics, 5071 West Saanich Rd, Victoria, BC V9E 2E7, Canada \\
$^7$Department of Physics and Astronomy, University of Victoria, Victoria, BC V8P 1A1, Canada \\
$^8$Laboratoire AIM CEA/DSM-CNRS-Universit´e Paris Diderot, IRFU/Service d’Astrophysique, CEA Saclay, F-91191 Gif-sur-Yvette, France \\
$^9$Department of Physics and Astronomy, University of Waterloo, Waterloo, Ontario N2L 3G1, Canada \\
$^{10}$East Asian Observatory, 660 N. A`oh\={o}k\={u} Place, University Park, Hilo, Hawaii 96720, USA 
$^{11}$LSST Project Office, 933 N. Cherry Ave, Tucson, AZ 85719, USA \\
$^{12}$Leiden Observatory, Leiden University, PO Box 9513, 2300 RA Leiden, The Netherlands
$^{13}$Max-Planck Institute for Astronomy, K{\"o}nigstuhl 17, 69117 Heidelberg, Germany \\
$^{14}$Max-Planck-Institut f{\"u}r extraterrestrische Physik, Giessenbachstrasse 1, 85748 Garching, Germany
$^{15}$Department of Physical Sciences, The Open University, Milton Keynes MK7 6AA, UK \\
$^{16}$The Rutherford Appleton Laboratory, Chilton, Didcot OX11 0NL, UK \\
$^{17}$Joint ALMA Observatory, Alonso de Córdova 3107, Vitacura - Santiago, Chile \\
$^{18}$Istituto di Astrofisica e Planetologia Spaziali-INAF, Via Fosso del Cavaliere 100, I-00133 Roma, Italy \\
}

\section*{APPENDIX A: HERSCHEL SOURCES}

\setcounter{figure}{0}
\setcounter{table}{0}

In this appendix we direct the reader to the full set of sources found in the filtered and unfiltered \emph{Herschel} 250-\um\ observations of the low-SCUBA-2-variance region of L1495 by CSAR. The unfiltered \emph{Herschel} sources are shown in Figures \ref{fig:finding_chart_A1} and \ref{fig:finding_chart_A2}, while the filtered \emph{Herschel} sources are shown in Figures \ref{fig:finding_chart_B1} and \ref{fig:finding_chart_B2}.  For the \emph{Herschel} sources, at DOI http://dx.doi.org/10.11570/16.0002, Table~A1 lists the name, right ascension and declination, measured major and minor FWHM sizes, position angle, flux densities as measured in \emph{Herschel} 160-\um, 250-\um, 350-\um\ and 500-\um\ emission, an extrapolated SCUBA-2 850-\um\ flux density (given in brackets), and the equivalent source in the \citet{marsh2016} catalogue.  Table~A2, at the DOI given above, lists the same properties for the filtered-\emph{Herschel} sources as given in Table~A1, except the 160-\um, 250-\um, 350-\um\ and 500-\um\ flux densities are measured in filtered Herschel emission, and the equivalent source in the \emph{Herschel} catalogue is listed.  The derived source properties: temperature, mass, density and deconvolved FWHM, are listed in Table~A3 at the DOI above for the \emph{Herschel} sources, and in Table~A4 at the DOI above for the filtered-\emph{Herschel} sources.

Many of the sources found in the unfiltered \emph{Herschel} data do not have counterparts in the \citet{marsh2016} catalogue.  \citet{marsh2016} identify sources using the \emph{getsources} algorithm \citep{menshchikov2012}, and produce two catalogues: an `unfiltered' catalogue, and a `filtered' catalogue of robustly detected cores, inclusion of a core into which requires a stringent set of criteria to be met \citep{marsh2016}.  The \citet{marsh2016} sources listed in the final column of Table~A1 are from their filtered catalogue.  We find that, when comparing our sources with the full \citet{marsh2016} catalogue, approximately half of the \emph{Herschel} sources without filtered \citet{marsh2016} counterparts can be identified with a source in the unfiltered \citet{marsh2016} catalogue (K. Marsh, priv. comm.).  The remainder of the sources are typically highly elongated and have a low signal-to-noise ratio, suggesting that with our chosen parameters for CSAR (optimised for detection of sources in the SCUBA-2 data), we may be detecting striations such as those identified by \citet{palmeirim2013}.  We therefore caution the reader that this is not the definitive Herchel catalogue for this region.  For the definitive Herschel catalogue, please see \citet{marsh2016}.

\clearpage

\renewcommand{\thefigure}{A\arabic{figure}a}
\begin{figure}
  \centering
   \includegraphics[width=0.5\textwidth]{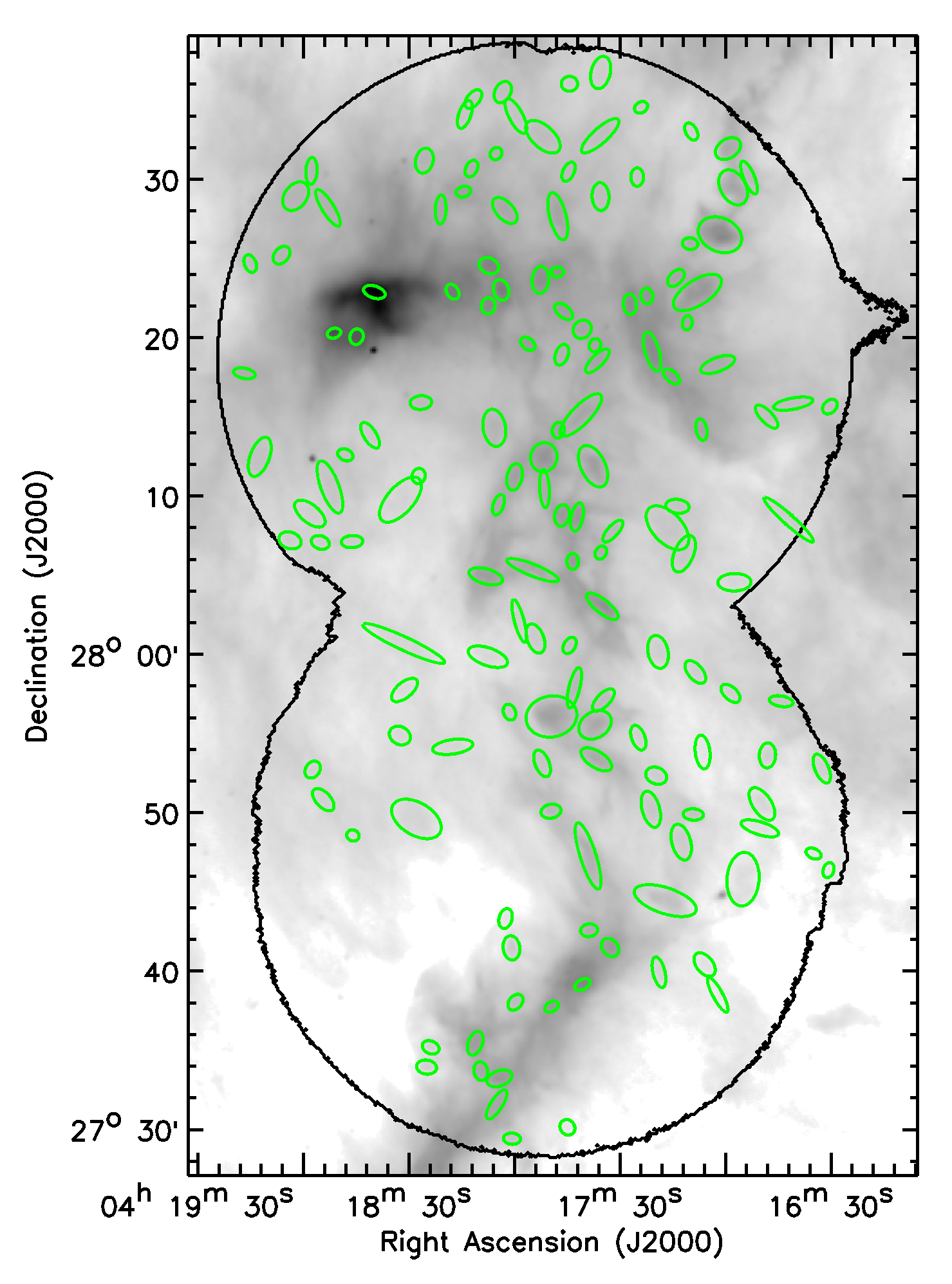}
  \caption{Grey-scale image of the head of the L1495 filament, in \emph{Herschel} 250\um\ emission. Sources detected in \emph{Herschel} 250\um\ emission are marked by small ellipses. The large-scale contour surrounds the region of lowest SCUBA-2 variance (c.f. \citealt{buckle2015}).}
  \label{fig:finding_chart_A1}
\end{figure}
\addtocounter{figure}{-1}
\renewcommand{\thefigure}{A\arabic{figure}b}
\begin{figure}
  \centering
  \includegraphics[width=0.5\textwidth]{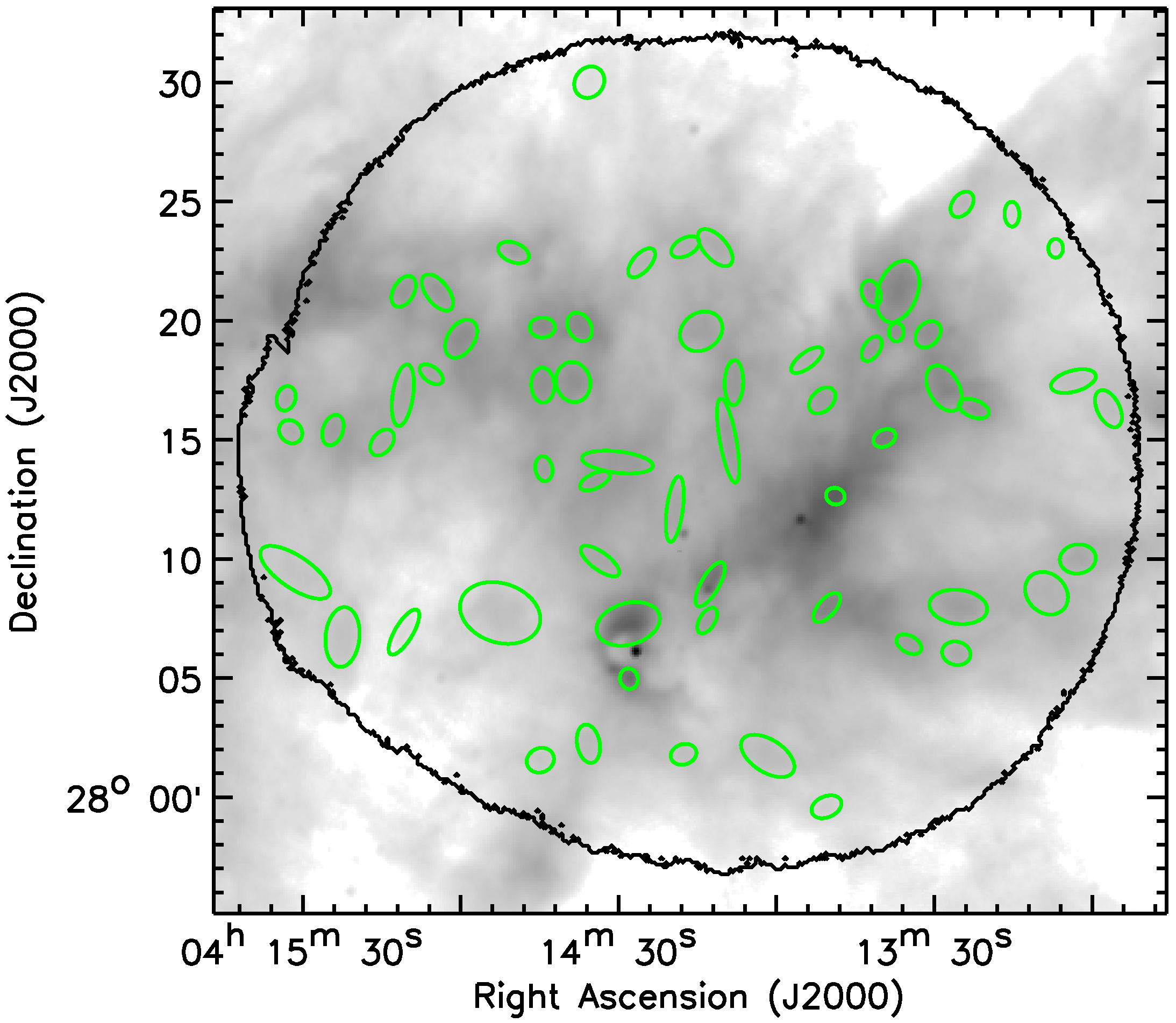}
  \caption{As Figure~\ref{fig:finding_chart_A1}, for the L1495 West region.}
  \label{fig:finding_chart_A2}
\end{figure}
\renewcommand{\thefigure}{\arabic{figure}}

\renewcommand{\thefigure}{A\arabic{figure}a}
\begin{figure}
  \centering
  \includegraphics[width=0.5\textwidth]{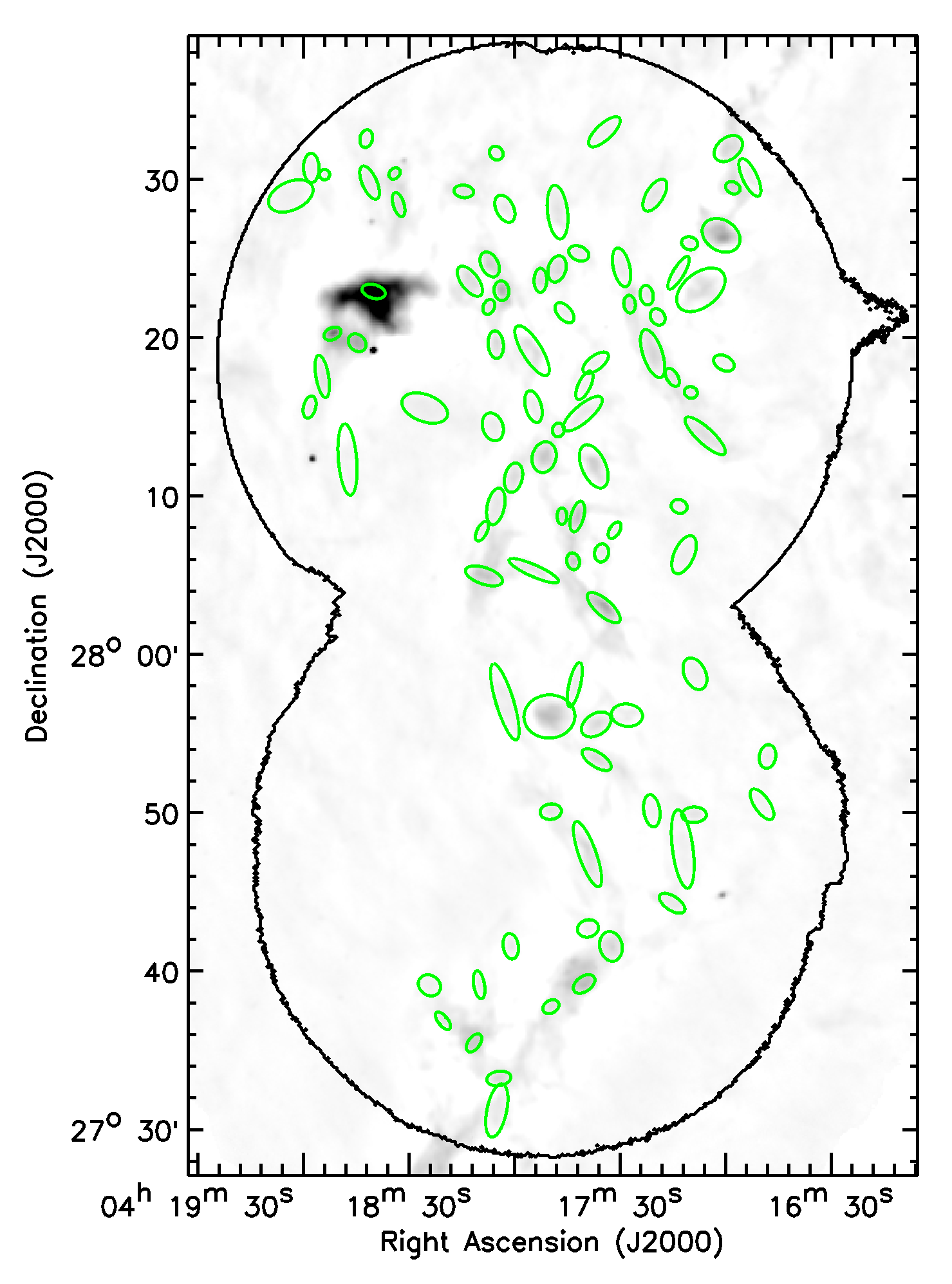}
  \caption{Grey-scale image of the head of the L1495 filament, in filtered \emph{Herschel} 250\um\ emission. Sources detected in filtered \emph{Herschel} 250\um\ emission are marked by small ellipses. The large-scale contour surrounds the region of lowest SCUBA-2 variance (c.f. \citealt{buckle2015}).}
  \label{fig:finding_chart_B1}
\end{figure}
\addtocounter{figure}{-1}
\renewcommand{\thefigure}{A\arabic{figure}b}
\begin{figure}
  \centering
  \includegraphics[width=0.5\textwidth]{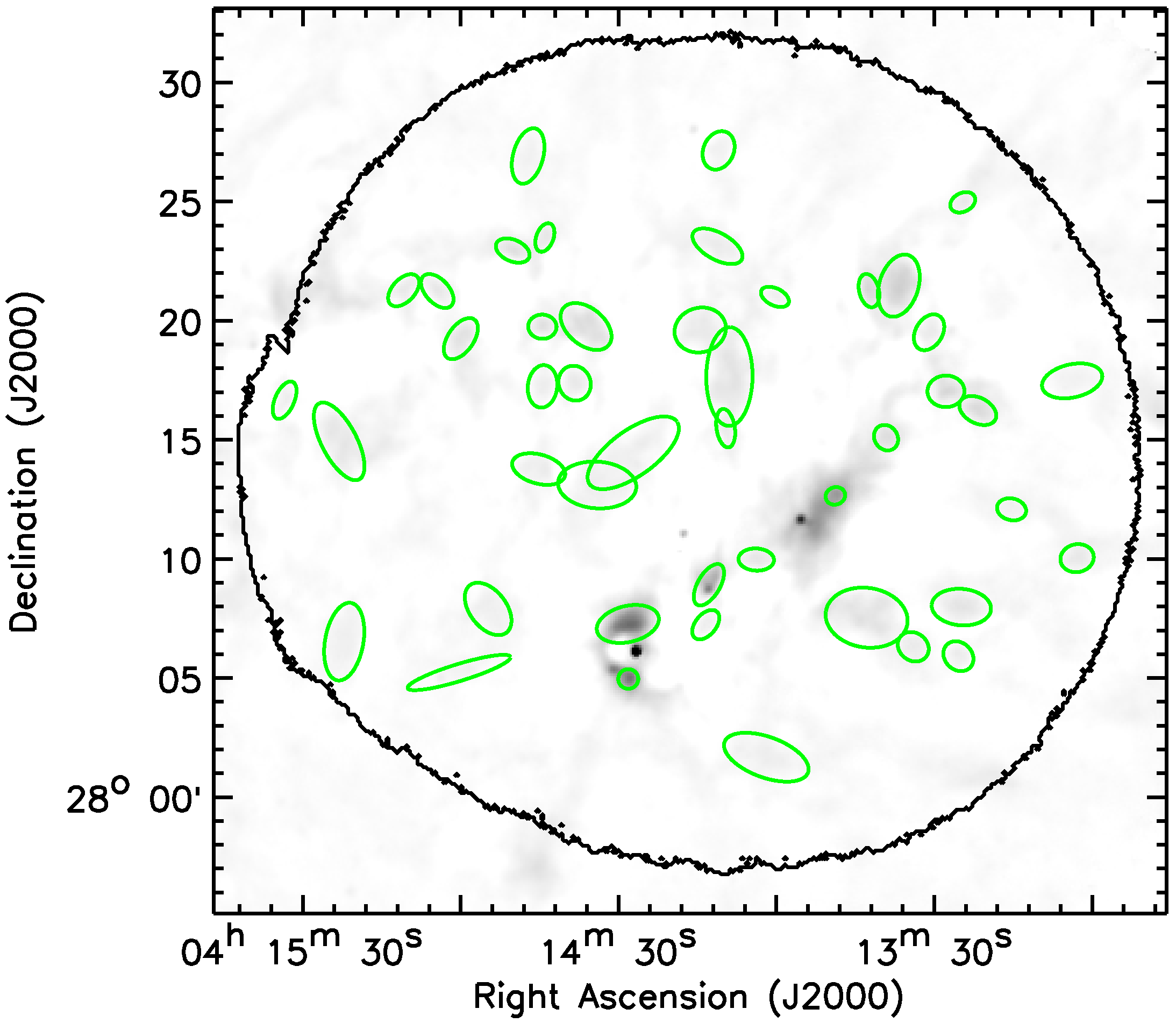}
  \caption{As Figure~\ref{fig:finding_chart_B1}, for the L1495 West region.}
  \label{fig:finding_chart_B2}
\end{figure}
\renewcommand{\thefigure}{\arabic{figure}}

\clearpage

\section*{APPENDIX B: DEMONSTRATION OF SURFACE BRIGHTNESS EFFECT}

\setcounter{figure}{0}
\setcounter{table}{0}

In this appendix we demonstrate that the increased number of sources detected in the spatially-filtered \emph{Herschel} 250-\um\ data compared to the number detected in the 850-\um\ SCUBA-2 data is due to the black-body function peaking at $\sim 250$\um, while 850\um\ is on the Rayleigh-Jeans tail, at temperatures typically found in molecular clouds.  Hence, dust at temperatures typically found in starless cores emits more flux at \emph{Herschel} wavelengths than at SCUBA-2 850-\um\ wavelength, because \emph{Herschel} wavelengths coincide with the peak of the black-body function.  For the same absolute noise level, any given core will appear at a higher signal-to-noise ratio to \emph{Herschel} than to SCUBA-2.

To test the hypothesis that this effect causes the increased detection of sources in the filtered \emph{Herschel} 250-\um\ data over the SCUBA-2 850-\um\ data, we scaled the noise on the \emph{Herschel} data such that the relative, rather than the absolute, sensitivity of the the \emph{Herschel} and SCUBA-2 maps became comparable.  In order to as closely mimic the SCUBA-2 data as possible, we added noise to the unfiltered \emph{Herschel} data before passing the data through the SCUBA-2 pipeline.

To do this, we determined the relative brightness of the 250-\um\ and 850-\um\ points on the grey-body curve, for a temperature $T=11.3$\,K (the mean temperature of our cores), and for a dust opacity index $\beta=1.3$ (as we use in the analysis in this work), i.e.
\begin{equation}
\frac{F_{250}}{F_{850}}=\left(\frac{\nu_{250}}{\nu_{850}}\right)^{3+\beta}\frac{e^{\frac{h\nu_{850}}{k_{\textsc{b}}T}}-1}{e^{\frac{h\nu_{250}}{k_{\textsc{b}}T}}-1}=4.11.
\end{equation}
We measured the RMS noise level of the SCUBA-2 850-\um\ map to be 0.9 mJy/6-arcsec pixel.  Hence, our target RMS value for the \emph{Herschel} data was $4.11\times0.9$ mJy/6-arcsec pixel, i.e. 3.7 mJy/6-arcsec pixel.  We determined the RMS noise level that needed to be added to the original map by subtracting the measured RMS noise level from the target noise level in quadrature.  We created a Gaussian noise distribution of the required amplitude, and added this to the unfiltered \emph{Herschel} 250-\um\ map.

We then passed the increased-noise \emph{Herschel} 250-\um\ map through the SCUBA-2 pipeline in the manner described in Section~\ref{sec:s2_herschel}, removing the large-scale structure from the \emph{Herschel} data.

We extracted sources from our increased-noise filtered \emph{Herschel} 250-\um\ map with CSAR, using the parameters described in Section~\ref{sec:s2_herschel}.  We extracted 47 sources from the region of lowest SCUBA-2 variance, of which 5 were excluded as being point sources.  This left 42 detections of sources likely to be starless cores.  We detected 25 starless cores in the 850-\um\ SCUBA-2 data.  While we detect more cores in the increased-noise filtered \emph{Herschel} 250-\um\ map, the cores we identify are similar to those detected in SCUBA-2 data: they are located along the dense filaments, and many can be directly associated with sources in the SCUBA-2 catalogue.  Particularly, the low-surface-brightness sources away from the dense filaments are no longer detected, as expected.   The starless cores which we detect in the increased-noise filtered \emph{Herschel} 250-\um\ map are listed in Table~\ref{tab:fakenoise_sources} and shown in Figures~\ref{fig:finding_chart_C1} and \ref{fig:finding_chart_C2}.

\renewcommand{\thefigure}{B\arabic{figure}a}
\begin{figure}
  \centering
  \includegraphics[width=0.5\textwidth]{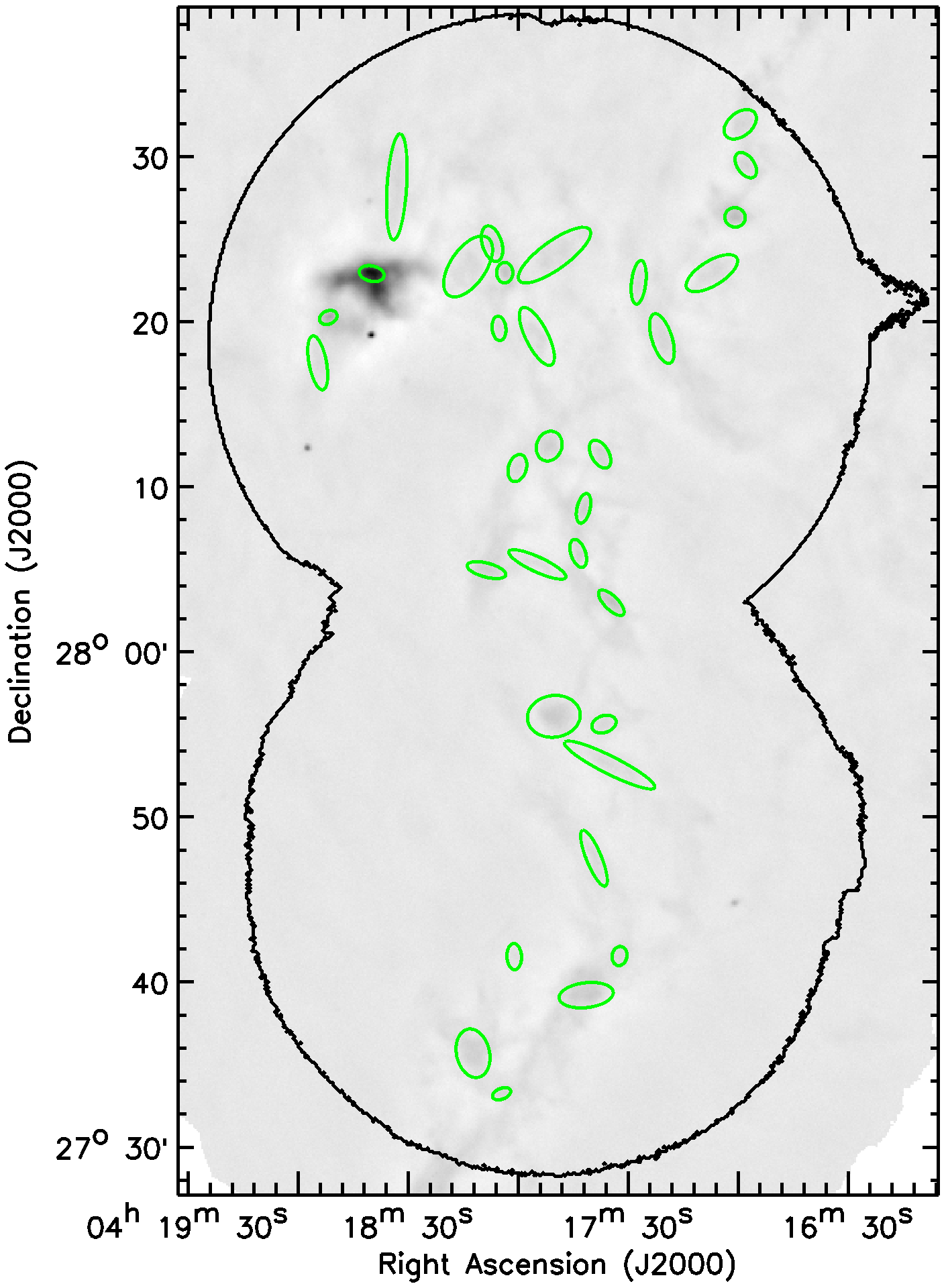}
  \caption{Grey-scale image of the head of the L1495 filament, in increased-noise filtered \emph{Herschel} 250\um\ emission. Sources detected in increased-noise filtered \emph{Herschel} 250\um\ emission are marked by small ellipses. The large-scale contour surrounds the region of lowest SCUBA-2 variance (c.f. \citealt{buckle2015}).  This should be compared with Figure~\ref{fig:finding_chart1}.}
  \label{fig:finding_chart_C1}
\addtocounter{figure}{-1}
\renewcommand{\thefigure}{B\arabic{figure}b}
  \includegraphics[width=0.5\textwidth]{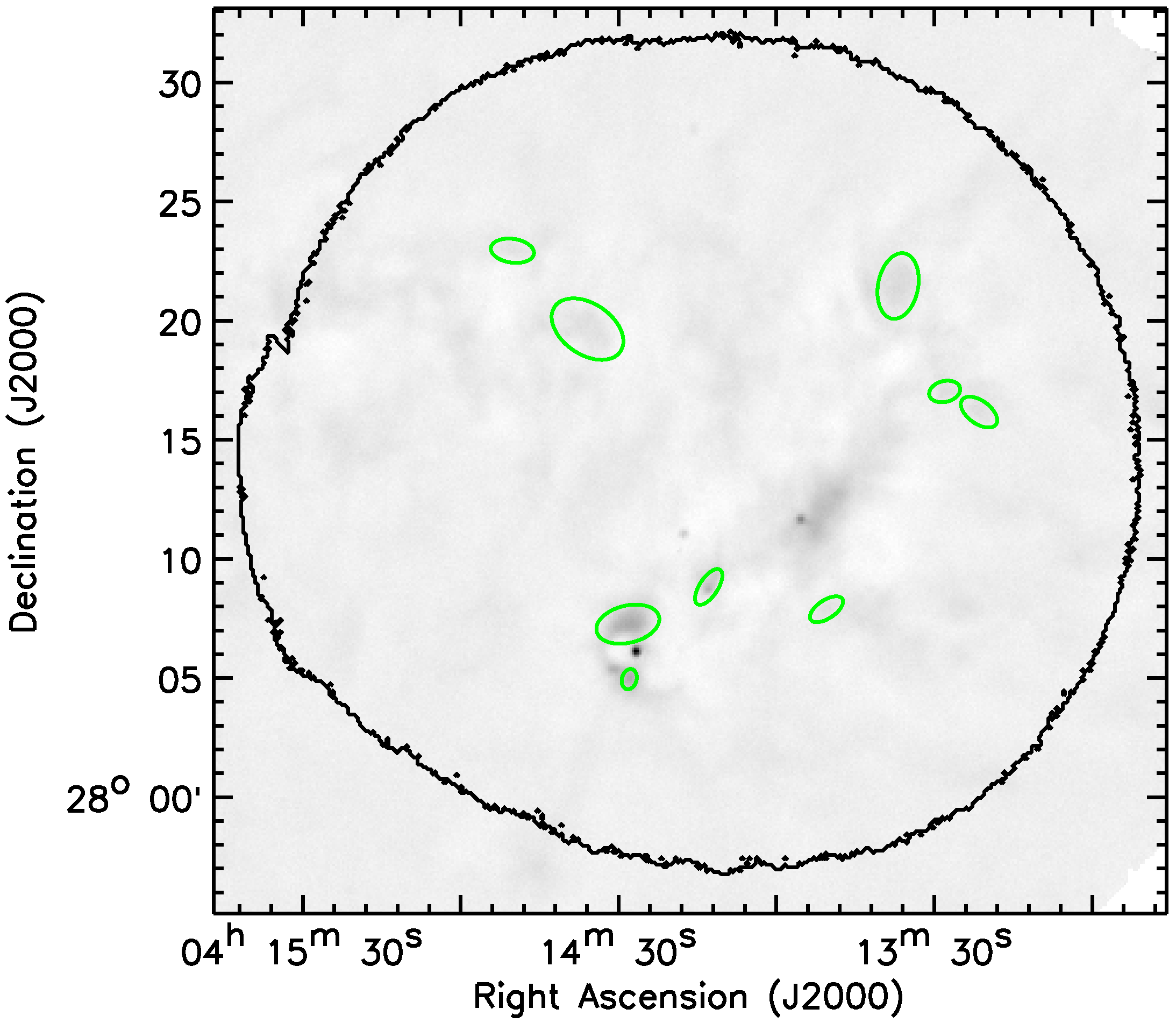}
  \caption{As Figure~\ref{fig:finding_chart_C1}, for the L1495 West region.  This should be compared with Figure~\ref{fig:finding_chart2}.}
  \label{fig:finding_chart_C2}
\end{figure}
\renewcommand{\thefigure}{B\arabic{figure}}

\renewcommand{\thetable}{B\arabic{table}}
\setlength{\tabcolsep}{2pt}
\begin{table}
\centering
\caption{Sources detected in increased-noise filtered \emph{Herschel} 250-\um\ data}
\label{tab:fakenoise_sources}
\begin{tabular}{c c c c c c}
\hline
Source & R.A. & Dec. & FWHM & Angle & Counterpart \\
Index & (J2000) & (J2000) & (arcsec) & ($^{\circ}$ E of N) & Sources \\
\hline
N1 & 4:18:40.53 & +28:23:02.2 & 7.7$\times$4.6 & 77.4 & S1,F1 \\
N2 & 4:14:27.71 & +28:07:12.8 & 13.5$\times$7.8 & 103.5 & S10,F3 \\
N3 & 4:18:52.44 & +28:20:22.1 & 5.5$\times$4.0 & 113.0 & S23,F2 \\
N4 & 4:14:27.50 & +28:04:53.9 & 4.3$\times$3.1 & 164.8 & S13,F4 \\
N5 & 4:14:12.23 & +28:08:45.4 & 8.6$\times$3.9 & 146.8 & S15,F5 \\
N6 & 4:17:00.29 & +28:26:28.2 & 6.2$\times$5.9 & 80.0 & S8,F6 \\
N7 & 4:17:34.40 & +28:03:05.5 & 10.2$\times$4.1 & 45.3 & S12,F8 \\
N8 & 4:18:03.83 & +28:23:06.4 & 6.2$\times$4.9 & 179.9 & S6,F10 \\
N9 & 4:17:50.17 & +27:56:11.8 & 16.1$\times$12.6 & 97.6 & S3,F12 \\
N10 & 4:17:42.04 & +28:08:48.8 & 9.2$\times$4.0 & 166.8 & S2,F13 \\
N11 & 4:17:43.49 & +28:06:05.0 & 8.8$\times$4.4 & 18.7 & S7,F15 \\
N12 & 4:17:41.24 & +27:39:17.2 & 16.5$\times$7.5 & 96.6 & S18,F11 \\
N13 & 4:18:12.20 & +27:35:45.1 & 14.9$\times$10.0 & 12.8 & S9,F17 \\
N14 & 4:18:08.72 & +28:05:03.8 & 12.1$\times$4.3 & 75.5 & S5,F14 \\
N15 & 4:17:51.53 & +28:12:35.5 & 9.2$\times$7.6 & 158.3 & S14,F16 \\
N16 & 4:17:37.51 & +28:12:06.2 & 9.3$\times$5.3 & 29.5 & S21,F19 \\
N17 & 4:18:07.34 & +28:24:51.8 & 11.1$\times$5.9 & 17.0 & S24,F18 \\
N18 & 4:16:57.33 & +28:29:38.3 & 8.7$\times$5.2 & 35.4 & S17,F20 \\
N19 & 4:18:04.37 & +27:33:18.0 & 5.8$\times$3.3 & 111.7 & F25 \\
N20 & 4:17:39.12 & +27:47:35.5 & 18.2$\times$4.5 & 22.3 & F26 \\
N21 & 4:16:58.84 & +28:32:06.9 & 11.2$\times$7.1 & 128.7 & F27 \\
N22 & 4:18:33.65 & +28:28:18.3 & 32.2$\times$5.9 & 176.3 & F29 \\
N23 & 4:13:35.62 & +28:21:21.9 & 14.0$\times$8.4 & 169.1 & F21 \\
N24 & 4:17:26.89 & +28:22:31.8 & 13.3$\times$4.3 & 173.4 & F28 \\
N25 & 4:18:05.39 & +28:19:43.5 & 7.5$\times$4.2 & 6.2 & F37 \\
N26 & 4:18:00.27 & +28:11:15.6 & 8.4$\times$5.4 & 163.5 & S19,F39 \\
N27 & 4:17:20.48 & +28:19:07.2 & 15.7$\times$6.1 & 17.7 & S25,F24 \\
N28 & 4:18:55.31 & +28:17:36.8 & 16.7$\times$5.3 & 10.8 & F44 \\
N29 & 4:17:32.10 & +27:41:39.7 & 5.8$\times$4.5 & 170.1 & F22 \\
N30 & 4:13:26.87 & +28:16:54.6 & 6.7$\times$4.4 & 104.1 & F34 \\
N31 & 4:14:35.08 & +28:19:38.3 & 16.7$\times$10.7 & 55.8 & F45 \\
N32 & 4:17:34.83 & +27:53:14.9 & 30.6$\times$5.4 & 63.3 & F35 \\
N33 & 4:17:06.70 & +28:23:05.1 & 17.7$\times$7.5 & 121.5 & F33 \\
N34 & 4:17:50.11 & +28:24:10.5 & 26.4$\times$8.5 & 125.8 & F40 \\
N35 & 4:17:36.40 & +27:55:43.6 & 7.6$\times$5.1 & 108.4 & F36 \\
N36 & 4:13:20.35 & +28:16:01.7 & 8.8$\times$4.8 & 53.8 & F41 \\
N37 & 4:18:00.96 & +27:41:37.3 & 7.9$\times$4.4 & 2.9 & F72 \\
N38 & 4:13:49.76 & +28:07:47.1 & 8.0$\times$3.8 & 123.5 & None \\
N39 & 4:17:54.95 & +28:19:14.6 & 19.4$\times$6.7 & 26.9 & F43 \\
N40 & 4:18:13.93 & +28:23:28.7 & 21.1$\times$10.8 & 145.8 & F38 \\
N41 & 4:17:54.74 & +28:05:24.7 & 19.0$\times$4.5 & 65.3 & F53 \\
N42 & 4:14:49.36 & +28:22:57.1 & 9.1$\times$5.0 & 82.0 & F49 \\
\hline
\end{tabular}
\end{table}
\setlength{\tabcolsep}{6pt}

\begin{figure}
  \centering
  \includegraphics[width=0.5\textwidth]{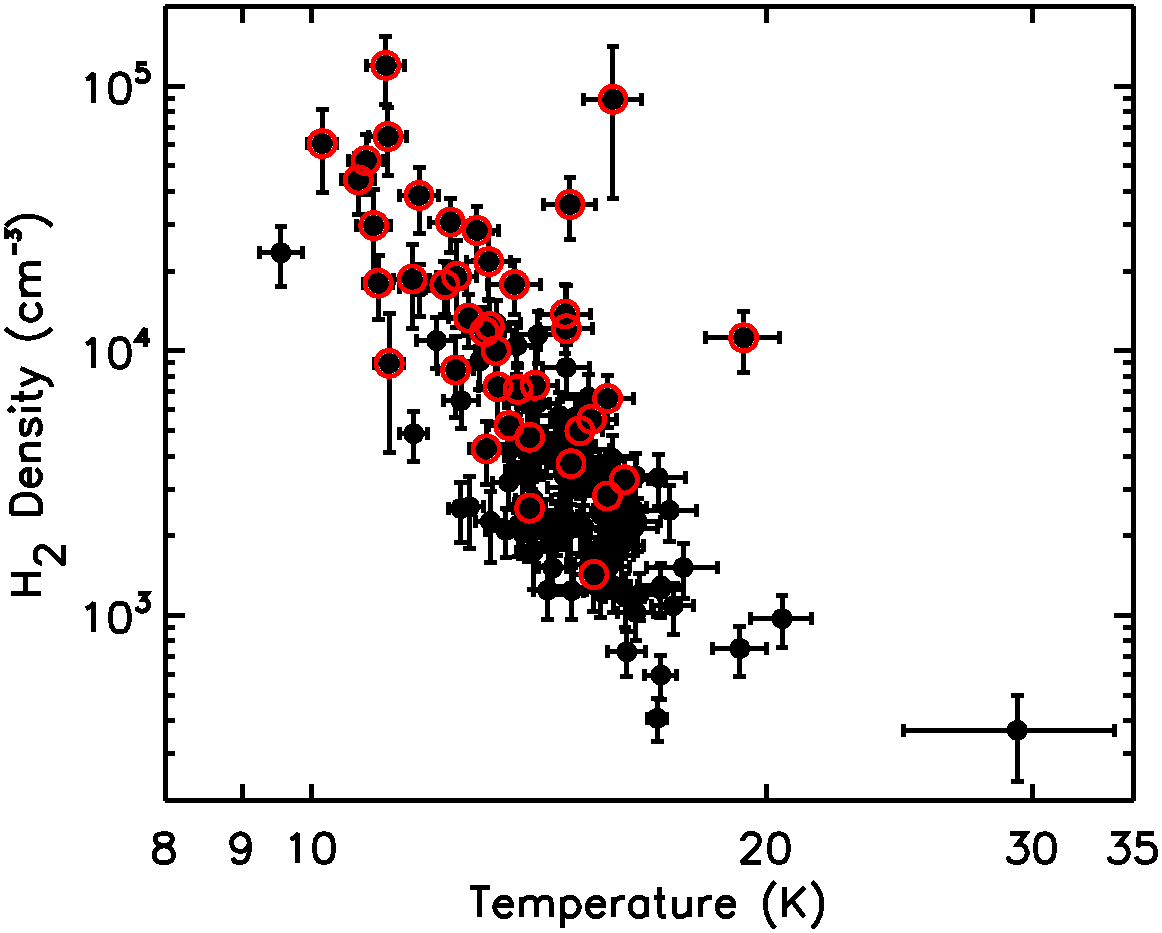}
  \caption{A comparison of the temperature and density of the filtered-\emph{Herschel} 250-\um\ sources, with those sources which are detected in the increased-noise filtered \emph{Herschel} data marked with a red circle.  Note the tendency for the sources detected in the increased-noise data to be among the denser sources.}
  \label{fig:fakenoise_temp_density}
\end{figure}

In Figure~\ref{fig:fakenoise_temp_density} we plot the temperatures and densities of the filtered-\emph{Herschel} 250-\um\ sources as determined in Section~\ref{sec:s2_herschel} (c.f. Figure~\ref{fig:temp_density}).  Those filtered-\emph{Herschel} sources which are detected in the increased-noise filtered \emph{Herschel} data (i.e. those sources which are listed as counterpart sources in Table~\ref{tab:fakenoise_sources}) are highlighted.  It can be seen that those sources detected in the increased-noise filtered \emph{Herschel} data are preferentially the denser sources.  The mean density of the filtered-\emph{Herschel} sources is $9.0\times10^{3}$\,H$_{2}$\,cm$^{-3}$, while the median density is $3.5\times10^{3}$\,H$_{2}$/cm$^{3}$.  The mean density of the subset of the filtered-\emph{Herschel} sources which are detected in the increased-noise data is $21.4\times10^{3}$\,H$_{2}$/cm$^{3}$, approximately 2.4 times that of the set as a whole, while the median density of the subset detected in the increased-noise data is $12.1\times10^{3}$\,H$_{2}$/cm$^{3}$, approximately 3.5 times that of the whole data set.  Moreover, 77\,per\,cent of the filtered-\emph{Herschel} sources with densities $\geq 10^{4}$\,H$_{2}$/cm$^{3}$ are detected in the increased-noise data, while only 16\,per\,cent of the filtered-\emph{Herschel} sources with densities $< 10^{4}$\,H$_{2}$/cm$^{3}$ are detected.

The clear cutoff in density which is apparent in the detection of SCUBA-2 850-\um\ sources is less distinct in the increased-noise filtered-\emph{Herschel} data.  This could be due to variation in temperature of the sources, meaning that our choice of increasing the RMS noise to a factor of 4.11 times the SCUBA-2 850-\um\ noise is not representative of all sources.  Resolution effects may also contribute to this.  However, these results support our hypothesis that the tendency for SCUBA-2 850-\um\ emission to detect the only densest sources is due to the decreased surface brightness of sources at 850-\um\ relative to 250-\um.

\end{document}